\documentclass[preprint,aps,pre]{revtex4}

\usepackage{epsfig}
\usepackage{psfig}

\newcommand{\erf}{{\rm erf}}
\newcommand{\be}{\begin{equation}}
\newcommand{\ee}{\end{equation}}
\newcommand{\ba}{\begin{eqnarray}}
\newcommand{\ea}{\end{eqnarray}}

\begin{document}

\title{A replica free evaluation of the neuronal population information with mixed 
continuous and discrete stimuli: from the linear to the asymptotic regime}

\author{Valeria Del Prete\thanks{vale@mth.kcl.ac.uk}}
 
\affiliation{Department of Mathematics, King's College London\\ 
Strand WC2R2LS, London, United Kingdom}

\bibliographystyle{unsrt}
\begin{abstract}
Recent studies have explored theoretically the ability of populations of neurons 
to carry information about a set of stimuli, both in the case of purely discrete
or purely continuous stimuli, and in the case of multidimensional 
continuous angular and discrete correlates, in presence of additional
quenched disorder in the distribution. An analytical expression for the mutual 
information has been obtained in the limit of large noise by means of the replica trick. 

Here we show that the same results can actually be obtained in most cases without 
the use of replicas,
by means of a much simpler expansion of the logarithm. Fitting the theoretical model to real neuronal 
data, we show that the introduction of correlations in the quenched disorder improves the fit, suggesting 
a possible role of signal correlations-actually detected in real data- in a redundant code.
We show that even in the more difficult analysis of the asymptotic regime, 
an explicit expression for the mutual 
information can be obtained without resorting to the replica trick despite the presence 
of quenched disorder, both with a gaussian and with a more realistic thresholded-gaussian model.
When the stimuli are mixed continuous and discrete, we find that 
with both models the information seem to grow logarithmically to infinity with the number of neurons and with the 
inverse of the noise, even though the exact general dependence cannot be derived explicitly for the thresholded gaussian 
model.
In the large noise limit lower values of information were obtained with the thresholded-gaussian model, for 
a fixed value of the noise and of the population size. On the contrary, in the asymptotic regime, with very low 
values of the noise, a lower information value is obtained with the gaussian model.
\end{abstract}

 
\maketitle 

\section{Introduction}
The mutual information, extensively used in the theory of communication \cite{Sha+48,Cov+91}, 
has been more recently proposed as a measure of the coding capacity of real neurons in the brain (see for example 
\cite{Rie+96,Rol+98,proc30}, for a general overview).
Information estimates, both from real data and in pure theoretical modelling, 
ideally quantify how efficiently an external observer might discriminate between 
several correlates of behaviour on the basis of the firing of single or multiple cells.

Several theoretical studies have explored 
the ability of one population of neurons to encode external stimuli, relevant to behaviour 
\cite{pap35,vale+01b,vale+01c,sompo+00,sompo+01}; others have tried to assess how efficiently 
the information is transmitted across several layers of a network, which may represent 
distinct stages of processing in some brain area \cite{Tre+92,Tre+95,Tre+98b}.

In most cited works the replica trick \cite{meza+87} has been successfully used in order to derive 
an explicit expression for the mutual information.
As we will show in detail in the next section, 
from the formula of the information replicas do appear as a natural 
methodological choice, due to the presence of the logarithm of a sum of conditional probabilities depending on some 
quenched parameters.
Yet in the cited works no attempt has been done to verify whether the same results can be obtained 
without resorting to replicas, even in the cases \cite{pap35,vale+01b,sompo+01} where the evaluation 
could be carried out without any additional 
assumption of replica symmetry.

Moreover an exact estimate of the mutual information regardless of the population size $N$ 
and of the noise $\sigma$
is often unachievable, so that an analytical 
expression can be provided only in some limit cases.
It might well be that restricting oneself to these cases makes the use of replicas 
redundant or at least an alternative choice to other methods.
 
In particular \cite{pap35,vale+01b} have used replicas to study the initial linear rise of the 
information, 
characterized by small population sizes and large noise in the firing 
distributions of the neurons; this limit would roughly correspond to a high 
temperature regime for a physical system
like a spin glass. It is reasonable to think that this limit can be treated and solved 
without replicas, since it is known that annealed and quenched averages coincide in the high 
temperature regime. 

Here we first reconsider the analysis performed in \cite{pap35,vale+01b}; we show that, 
in the limit when the noise 
$\sigma$ is large and the population size $N$ is small, the same analytical  
expressions for the information can be obtained without the use of the replica trick, 
by means of a simple Taylor expansion of the logarithm,  regardless of the nature of the stimulus whether 
purely discrete or mixed continuous and discrete, and both with a gaussian and with a more 
realistic thresholded-gaussian firing distribution.

In the particular case of mixed continuous angular and discrete stimuli \cite{vale+01b}, 
the distribution had been parameterized in order to model the firing of neurons 
recorded from the motor cortex of monkeys performing arm movements, categorized according to a direction and a "type" 
\cite{Don+98,vale}. 
Restricted to this data set, correlations in the preferred direction of a given unit across different movement types 
were actually observed, but the impact of such correlations on the information content was not quantified. 
Thus here we investigate theoretically whether correlations introduced in the quenched parameters characterizing the distribution 
can improve the fit of real information curves 
provided by the model. This would suggest that such correlations are information bearing, or better, depress the information, leading 
to a redundant code.

We move then to the limit of large population sizes and small noise.
An attempt to study this regime in the presence of purely discrete stimuli by means of replicas 
was unsuccessful in \cite{pap35}.

Here we show that even in the asymptotic regime, in the case of purely discrete stimuli, an analytical expression 
for the mutual information can be provided without the use of the replica trick.

Another replica free approach to this limit was proposed in \cite{sompo+00}, applicable both to the case of 
continuous and discrete stimuli, for a generic firing distribution, provided that it can be factorized into single neuron 
probability density functions.
No additional quenched disorder was assumed in the distribution.
Here we try and apply this method to our particular model and we find the assumption under which we retrieve our original approximation.

Finally, in \cite{vale+01b} it has been shown that, when limited to the initial linear regime, 
both the gaussian and the thresholded gaussian model 
provide the same analytical expression for the mutual information, except for renormalization 
of a noise parameter. In particular lower values of the information were obtained with the thresholded gaussian model.
We investigate this issue in the asymptotic regime comparing the leading term of the information for both models.

\section{Population information in the initial linear regime}
\subsection{Coding of purely discrete and mixed continuous and discrete stimuli 
in a gaussian approximation}\label{large_discr}

The firing of neurons emerging from the analysis of real data is characterized by 
strong irregularities and by a wide variability. The choice of a gaussian model 
as a possible firing rate distribution 
might therefore seem unrealistic and unjustified.
Yet with a large sample of data it is likely that most irregularities in the distribution 
average out; their presence if often due to a too poor sampling, which in turn biases information 
estimates, so that smoothing with a gaussian or other kernels has become a standard procedure 
in data analysis (see \cite{Rol+98} for a review of several regularizing procedures).
The advantage in using a gaussian approximation is easier mathematical analysis, which allows for
the derivation of an explicit expression for the mutual information 
\cite{pap35,vale+01b,vale+01c,sompo+01}. 
Moreover, at least in the regime of the initial 
information rise, the use of a more realistic model leads to the same 
mathematical expression for the mutual information, 
except for a renormalization of the noise \cite{vale+01b}. This last issue will be discussed 
more in detail in the next section.

Let us consider a population of $N$ independent cells 
which fire to a set of $p$ discrete stimuli, parameterized by a discrete variable $s$, 
according to a gaussian distribution:
\begin{equation}
p(\{\eta_i\}|s)=\prod_{i=1}^N \frac{1}{\sqrt{2\pi\sigma^2}}
exp-\left[\left(\eta_i-\eta_i^s
\right)^2/2\sigma^2\right];
\label{dist}
\end{equation}
where $\eta_i$ is the firing rate  of the $i^{th}$ input neuron, while $\eta_i^s$ is its 
mean rate in response to stimulus $s$.

The mutual information between the neuronal firing rates $\{\eta_i\}$ and the stimuli 
$s$ reads:
\be
I(\{\eta_i\},s)=\sum_{s} p(s)
\int \prod_i d\eta_i
p(\{\eta_i\}|s)\log_2\frac{p(\{\eta_i\}|s)}{p(\{\eta_i\})}; 
\label{info} 
\ee

Since $p(\{\eta_i\}$ can be written as $\sum_s p(s) p(\{\eta_i\}|s)$ it is easy to show 
that the mutual information can be expressed as 
the difference between the entropy of the firing rates 
$H(\{\eta_i\})$ and the equivocation $\left\langle
H(\{\eta_i\}|s) \right\rangle_{s}$:
\begin{equation}
I(\{\eta_i\},s)=H(\{\eta_i\})-\left\langle
H(\{\eta_i\}|s) \right\rangle_{s}
\end{equation}
with:
\begin{equation}
\left\langle H(\{\eta_i\}|s)\right\rangle_{s}=-
\sum_{s} p(s)\int \prod_i d\eta_i 
p(\{\eta_i\}|s)\log_2
p(\{\eta_i\}|s);
\label{equiv}
\end{equation}
\begin{equation}
H(\{\eta_i\})=-\sum_{s}p(s)\int \prod_i d\eta_i 
p(\{\eta_i\}|s)\log_2\left[\sum_{s^\prime}p(s^\prime)
p(\{\eta_i\}|s^\prime)\right].
\label{outent}
\end{equation}

The variables $\{\eta_i\}$ in $p(\{\eta_i\}|s^\prime)$ 
are {\it quenched}: the sum on the stimuli $s^\prime$ should be performed, 
and the logarithm
taken, for any fixed configuration $\{\eta_i\}$, before integrating on $\{\eta_i\}$.
The replica trick, devised to perform averages of the partition function across 
quenched disorder in spin glasses \cite{meza+87}, seems to apply also to this case.
Yet, 
contrary to what is found in the theory of spin glasses, where the connectivities 
vary on a much longer time scale with respect to the spins and therefore they 
are quenched, here the presence of 
quenched disorder does not reflect any real distinction between two separate time scales. 
In fact the same sum appears outside the logarithm, and if one were able to explicitly
derive  $p(\{\eta_i\})$ from $p(\{\eta_i\}|s^\prime)$ there would be no need for replicas 
to evaluate  $H(\{\eta_i\})$. 
 
In the specific case of the distribution (\ref{dist}) $p(\{\eta_i\})$ has a functional 
dependence on the configuration of the average rates $\{\eta_i^s\}$ and it cannot be 
explicitly derived except for some trivial cases, like:
\be
\eta_i^s=\eta^0_i\,\,\,\,\forall s;
\ee
where the information is obviously zero, since $p(\{\eta_i\}|s)$ does not depend on $s$ anymore;
or the opposite noiseless limit, where the cells fire at each stimulus $s$ always 
with a pattern $\{\eta_i^s\}$ and the configurations $\{\eta_i^s\}$ across the stimuli do not 
overlap. In this case, when the stimuli are equally likely, so that $p(s)=1/p$ one has:
\be
p(\{\eta_i\}|s)=\delta(\{\eta_i\}-\{\eta_i^s\});
\ee
\be
p(\{\eta_i\})=1/p\sum_{s^\prime}\delta(\{\eta_i\}-\{\eta_i^{s^\prime}\});
\ee
and since the average configurations $\{\eta_i^s\}$ do not overlap 
it is easy to see that the mutual information reaches the upper bound of $\log_2 p$.

In a more realistic context, the average firing rates $\{\eta_i^s\}$ are not kept fixed, 
reflecting the strong variability of the neural activity detected in real data.
Therefore, 
in order to obtain an information estimate independent of a particular 
configuration of the selectivities, the variables 
$\{\eta_i^s\}$ are considered quenched
and the information must finally be averaged across the distribution of $\{\eta_i^s\}$:
\be
I(\{\eta_i^s\})\longrightarrow\left\langle I(\{\eta_i^s\})\right\rangle_\eta  
\ee
\be
\left\langle F(\{\eta_i^s\})\right\rangle_\eta=\int d\{\eta_i^s\} \varrho(\{\eta_i^s\})
F(\{\eta_i^s\});
\ee
$I(\{\eta_i^s\})$ is the mutual information between the neuronal 
firing rates $\{\eta_i\}$ and the stimuli $s$ evaluated according to eq.(\ref{info}), for 
a particular configuration of the mean rates $\{\eta_i^s\}$.

This approach has been followed in \cite{pap35,vale+01b}, where the replica trick 
has been used to perform the analytical evaluation.

Let us consider the case where quenched disorder is uncorrelated 
and identically distributed
across units and across the $p$ discrete correlates:
\begin{equation}
\varrho(\{\eta^i_s\})=\prod_{i,s}\varrho(\eta_s^i)=\left[\varrho(\varepsilon)\right]^{Np}
\label{ro_eps}
\end{equation}
 
As already shown in \cite{pap35,vale+01b}, it is easy to prove that for a population 
of independent units the equivocation $\left\langle H(\{\eta_i\}|s)\right\rangle_s$ 
is additive. 

By $I(\{\eta_i\},s)$ and 
$H(\{\eta_i\})$ and $\left\langle
H(\{\eta_i\}|s) \right\rangle_{s}$ in the following I will implicitly 
mean the corresponding quenched averaged quantities. 
Inserting eq.(\ref{dist}) in eq.(\ref{equiv}) one obtains:
\be
\left\langle H(\{\eta_i\}|s)\right\rangle_{s}=\frac{N}{2\ln 2}
\left(1+\ln 2\pi\sigma^2\right);
\label{final_equi}
\ee

I turn now to the more difficult evaluation of the rate entropy. 
Inserting eq.(\ref{dist}) in eq.(\ref{outent}) and using the equivalence:
\be
\ln\left[\sum_{s^\prime} p(s^\prime) 
\prod_i \frac{e^{-(\eta_i-\eta_i^{s^\prime})^2/2\sigma^2}}
{\sqrt{2\pi\sigma^2}}\right]=
-\frac{N}{2}\ln 2\pi\sigma^2-\sum_i \frac{\eta_i^2}{2\sigma^2}
+\ln\left[\sum_{s^\prime} p(s^\prime) 
\prod_i e^{(2\eta_i\eta_i^{s^\prime}
-(\eta_i^{s^\prime})^2)/2\sigma^2}\right];
\label{ch_var}
\ee
one can integrate out the second term on the rhs in eq.(\ref{ch_var}); the result,  
added to the first term, simplifies with the equivocation, eq.(\ref{final_equi}); 
rearranging all the terms  the mutual information can be written in the following form:
\be
I(\{\eta_i\},s)=-\left\langle \frac{1}{\ln 2} \sum_{s}p(s)\int \prod_i d\eta_i \prod_i 
e^{-\left(\eta_i-\eta_i^s\right)^2/2\sigma^2}
\ln\left[\sum_{s^\prime} p(s^\prime) 
\prod_i e^{-((\eta_i^s)^2+(\eta_i^{s^\prime})^2
-2\eta_i\eta_i^{s^\prime})/2\sigma^2}\right]\right\rangle_\eta;
\label{no_appr}
\ee
 
Due to the presence of the sums and of the quenched disorder under the logarithm 
an analytical expression of the mutual information  
cannot be obtained in the general case; 
yet the evaluation can be performed in some limit cases.
We focus here on the initial regime, where the number of cells is not large compared to the 
noise. The asymptotic regime for large  population sizes will be discussed later on.

As it has been shown in \cite{pap35}, one way to get rid of the logarithm and to perform 
the quenched averages and the sums in eq.(\ref{no_appr}) is by means of the replica trick 
\cite{meza+87}; yet, in the limit when the noise $\sigma$ is very large and the population 
size $N$ is not large, a straightforward and natural approach consists in performing 
a simple taylor expansion of the exponentials under the logarithm:

\ba
&&\ln\left[\sum_{s^\prime} p(s^\prime) 
\prod_i e^{-((\eta_i^s)^2+(\eta_i^{s^\prime})^2
-2\eta_i\eta_i^{s^\prime})/2\sigma^2}\right]\nonumber\\
&&\simeq
\ln\left[1-\sum_{s^\prime} p(s^\prime)\left(\sum_i \frac{(\eta_i^{s})^2}{2\sigma^2}
+\sum_i\frac{(\eta_i^{s^\prime})^2}{2\sigma^2}-\sum_i\frac{\eta_i\eta_i^{s^\prime}}
{\sigma^2}-\frac{1}{2}\sum_{ij}\frac{\eta_i\eta_j\eta_i^{s^\prime}
\eta_j^{s^\prime}}{\sigma^4}\right)\right]\nonumber\\
&&\simeq -\sum_{s^\prime} p(s^\prime)\left(\sum_i \frac{\left(\eta_i^{s}\right)^2}{2\sigma^2}
+\sum_i\frac{(\eta_i^{s^\prime})^2}{2\sigma^2}-\sum_i\frac{\eta_i\eta_i^{s^\prime}}
{\sigma^2}-\frac{1}{2}\sum_{ij}\frac{\eta_i\eta_j\eta_i^{s^\prime}
\eta_j^{s^\prime}}{\sigma^4}\right)\nonumber\\
&&-
\frac{1}{2}\sum_{s^\prime} p(s^\prime)\sum_{s^{\prime\prime}} p(s^{\prime\prime})
\sum_{ij}\frac{\eta_i\eta_j\eta_i^{s^\prime}
\eta_j^{s^{\prime\prime}}}{\sigma^4};
\label{log_exp}
\ea

The terms order $1/\sigma^4$ must be kept because it can be shown that after integration 
on $\{\eta_i\}$ they will actually give a contribution order $1/\sigma^2$ to the mutual information.

Inserting the expansion (\ref{log_exp}) in eq.(\ref{no_appr}) and performing the integration 
on $\{\eta_i\}$ one obtains:
\ba
&&I(\{\eta_i\},s)=\frac{1}{\ln 2}\frac{1}{2\sigma^2}\left\langle\sum_i
\left[\sum_s p(s)\left(\eta_i^s\right)^2-\sum_s\sum_{s^\prime} p(s) p(s^\prime)
\eta_i^s\eta_i^{s^\prime}\right]\right\rangle_\eta\nonumber\\
&&=\frac{1}{\ln 2}\frac{N}{2\sigma^2}
\sum_s\sum_{s^\prime} p(s) p(s^\prime)\left[\left\langle\left(\eta^s\right)^2
\right\rangle_\eta-\left\langle\eta^s\eta^{s^\prime}\right\rangle_\eta\right];
\label{log_exp_ins}
\ea

where we have used the fact that quenched disorder is uncorrelated and identically distributed 
across stimuli and neurons.

The averages across quenched disorder and across $s$,$s^\prime$ can be performed distinguishing 
between the cases $s=s^\prime$,$s\neq s^\prime$.
The final result for the mutual information up to order $N/\sigma^2$ reads:
\be
I(\{\eta_i\},s)=\frac{1}{\ln 2}\frac{N}{2\sigma^2}\frac{p-1}{p}\sigma^2_\eta;
\,\,\,\,\,\,\sigma^2_\eta=
\left\langle\eta^2\right\rangle_\eta-\left\langle\eta\right\rangle^2_\eta;
\ee
The same result has been obtained in \cite{pap35} by means of the replica trick.
We have checked that the agreement between the two approaches is found also at higher 
orders in $N/\sigma^2$.
Yet the derivation via the replica trick is clearly longer and more complicated, 
and a priori less controllable than the simple Taylor expansion used here to derive 
the same results.

The interest in the coding of purely discrete stimuli rises naturally from the need to 
provide a theoretical framework allowing a direct quantitative comparison with the results 
of real experiments. In fact
in a typical experimental protocol neural activity is recorded from some 
areas in the brain, while the subject (human or animal) is presented a discrete 
number of stimuli, 
or it is trained to perform a discrete number of tasks.

Yet natural stimuli are multi-dimensional and some of the dimensions can vary 
in a continuous domain. For example a visual stimulus can be parameterized through 
its colour (varying within a discrete set of possible choices) and its orientation 
(represented by continuous angle).
It is therefore a primary theoretical interest to extend our results to the case where 
the stimulus may is multi-dimensional and the dimensions may be discrete and continuous.

In \cite{vale,vale+01b,vale+01c} the coding of movements categorized according to their 
direction (continuous dimension) and their {\it type} (discrete dimension) has been 
studied via direct information estimates from real data and pure theoretical modelling.
In particular in \cite{vale+01b} the information between the neuronal firing rates and the 
movements has been evaluated in the limit of large noise and finite population size, 
in presence of quenched disorder and resorting to the replica trick.

In analogy to the model studied in \cite{vale+01b} let us consider 
a population of $N$ neurons firing independently of one another 
to an external stimulus 
parameterized by an angle $\vartheta$ and a discrete 
variable $s$, according to a gaussian distribution:
\begin{equation}
p(\{\eta_i\}|\vartheta,s)=\prod_{i=1}^N \frac{1}{\sqrt{2\pi\sigma^2}}
exp-\left[\left(\eta_i-\tilde{\eta}_i(\vartheta,s)
\right)^2/2\sigma^2\right];
\label{dist2}
\end{equation}
Like in eq.(\ref{dist}), $\eta_i$ is the firing rate of the $i^{th}$ neuron;
$\tilde{\eta}_j(\vartheta,s)$ is its average firing rate corresponding to the stimulus 
$(\vartheta,s)$:

\begin{equation}
\tilde{\eta}_i(\vartheta,s)=\varepsilon^i_s\bar{\eta}_i(\vartheta)+(1-\varepsilon_s^i)
\eta^f;
\label{tuning_tot}
\end{equation}
\begin{equation}
\bar{\eta}_i(\vartheta-\vartheta^0_{i,s})=
\eta^0
\cos^{2m}\left(\frac{\vartheta-\vartheta^0_{i,s}}{2}\right);
\label{tuning2}
\end{equation}
where $\varepsilon_s^i$ and $\vartheta^0_{i,s}$ are sources of quenched disorder, 
distributed respectively between $0$ and $1$ and between $0$ and $2\pi$ and I assume that 
quenched disorder is uncorrelated and identically distributed across neurons and stimuli:

\begin{equation}
\varrho(\{\varepsilon^i_s\})=\prod_{i,s}\varrho(\varepsilon_s^i)=
\left[\varrho(\varepsilon)\right]^{Np}
\label{ro_eps2}
\end{equation}
\be
\varrho(\{\vartheta^0_{i,s}\})=\left[\varrho(\vartheta^0)\right]^{Np}=\frac{1}{(2\pi)^{Np}}
\label{ro_0}
\ee

Eq.(\ref{tuning_tot}) states that for each discrete correlate $s$ each neuron $i$ fires 
at an average rate modulating with $\vartheta$ around the preferred direction 
$\vartheta^0_{i,s}$ with an amplitude $\varepsilon_s^i$; alternatively the 
average rate is fixed to a value $\eta^f$, 
independently of $\vartheta$, with amplitude $1-\varepsilon_s^i$.
In \cite{vale+01b} it has been shown that a similar choice for the average rate can 
effectively reproduce the main features of real neurons directional tuning curves.

The basic definitions (\ref{info}),(\ref{equiv}),(\ref{outent}) as well as the initial treatment can 
be easily generalized to the case of structured stimuli, via the replacements:
\ba
\sum_s p(s)&\longrightarrow &\sum_s p(s)\int d\vartheta p(\vartheta)\nonumber\\
\eta_i^s &\longrightarrow & \tilde{\eta}_i(\vartheta,s)\nonumber\\
\left\langle..\right\rangle_\eta &\longrightarrow &\left\langle..\right\rangle_{\varepsilon,\vartheta^0}
\label{rep}
\ea
It is easy to show that the mutual 
information can be expressed in a form analogous to eq.(\ref{no_appr}):
\ba
&&I(\{\eta_i\},\vartheta\otimes s)=-\left\langle \frac{1}{\ln 2} \sum_{s}p(s)
\int d\vartheta p(\vartheta) \int \prod_i d\eta_i \prod_i 
e^{-\left(\eta_i-\tilde{\eta_i}(\vartheta,s)\right)^2/2\sigma^2}\right.\nonumber\\
&&
\left.\ln\left[\sum_{s^\prime} p(s^\prime) \int d\vartheta^\prime
\prod_i e^{-\left(\left(\tilde{\eta}_i(\vartheta,s)\right)^2+
\left(\tilde{\eta}_i\left(\vartheta^\prime,s^\prime\right)\right)^2
-2\eta_i\tilde{\eta}_i\left(\vartheta^\prime,s^\prime\right)\right)/2\sigma^2}\right]
\right\rangle_{\varepsilon,\vartheta^0};
\label{no_appr2}
\ea
 
We use again an expansion of the logarithm similar to (\ref{log_exp}); 
it is then easy to derive the 
analogous of eq.(\ref{log_exp_ins}):
\ba
&&I(\{\eta_i\},\vartheta\otimes s)\simeq\frac{1}{\ln 2}\frac{N}{2\sigma^2}
\sum_s\sum_{s^\prime} p(s) p(s^\prime)\int d\vartheta \int d\vartheta^\prime 
p(\vartheta) p(\vartheta^\prime) \left[\left\langle
\left(\tilde{\eta}(\vartheta,s)\right)^2
\right\rangle_{\varepsilon,\vartheta^0}-
\left\langle\tilde{\eta}(\vartheta,s)\tilde{\eta}(\vartheta^\prime,s^\prime)
\right\rangle_{\varepsilon,\vartheta^0}\right];\nonumber
\label{log_exp_ins2}
\ea

From eqs.(\ref{tuning_tot}),(\ref{tuning2}),(\ref{ro_eps2}),(\ref{ro_0}) 
it is easy to verify that:

\begin{eqnarray}
&&\sum_s\int d\vartheta p(s)p(\vartheta)
\langle[\tilde{\eta}(\vartheta,s)]^2\rangle_{\varepsilon,\vartheta^0}\nonumber\\
&&=(\eta^0)^2\left[(A_2+\alpha^2-2\alpha A_1)
\langle\varepsilon^2\rangle_{\varepsilon}+\alpha^2+2\alpha(A_1-\alpha)\langle\varepsilon
\rangle_{\varepsilon}\right];
\label{Lambda1}
\end{eqnarray}
\begin{eqnarray}
&&\sum_{s}\sum_{s^\prime}p(s)p(s^\prime)\int d\vartheta d\vartheta^\prime 
p(\vartheta)p(\vartheta^\prime)
\langle\tilde{\eta}(\vartheta,s)\tilde{\eta}(\vartheta^\prime,s^\prime)
\rangle_{\varepsilon,\vartheta^0}\nonumber\\
&&=(\eta^0)^2\left[(A_1-\alpha)^2\left(\frac{p-1}{p}
\langle\varepsilon\rangle_{\varepsilon}^2+\frac{1}{p}\langle\varepsilon^2
\rangle_{\varepsilon}\right)+\alpha^2+2\alpha(A_1-\alpha)
\langle\varepsilon\rangle_{\varepsilon}\right];
\label{Lambda2}
\end{eqnarray}
\begin{equation}
A_1=\frac{1}{2^{2m}}\left(\begin{array}{c}2m\\m\end{array}\right);\,\,\,
A_2=\frac{1}{2^{4m}}\left(\begin{array}{c}4m\\2m\end{array}\right);\,\,\,
\alpha=\frac{\eta^f}{\eta^0}.
\label{A_2}
\end{equation}

Inserting eqs.(\ref{Lambda1}),(\ref{Lambda2}) in eq.(\ref{log_exp_ins2}) 
one obtains the final expression for the mutual information up to order 
$N/\sigma^2$:

\begin{equation}
I(\{\eta_i\},\vartheta\otimes s)\simeq\frac{1}{\ln2}
\frac{N(\eta^0)^2}{4\sigma^2}
\left[\frac{p-1}{p}2\left(\alpha-A_1\right)^2\sigma^2_\varepsilon+2\left(A_2-(A_1)^2\right)
\left\langle\varepsilon^2\right\rangle_\varepsilon\right];
\label{final_info2}
\end{equation}

In \cite{vale+01b} it has been shown that the same expression for the information 
in linear approximation is obtained
in the case where the preferred directions do not modulate 
with the discrete stimuli: $\vartheta^0_{i,s}=\vartheta^0_{i}\,\,\forall s,i$.
It can be easily proved that a different contribution to the information
would derive in either case from the term 
$\left\langle\tilde{\eta}(\vartheta,s)\tilde{\eta}(\vartheta^\prime,s^\prime)
\right\rangle_{\varepsilon,\vartheta^0}$; yet the differing term becomes zero when averaged  
across $\vartheta$,$\vartheta^\prime$. 

Setting $\vartheta^0_{i,s}=\vartheta^0_{i}\,\,\forall s$ corresponds to correlating 
the signal that each neuron carries about different stimuli $s$. 
Intuitively, since no difference 
in the preferred orientation can be detected any more while looking at distinct 
correlates $s$, one would expect an information loss.
Such a loss is indeed present, as revealed from 
a detailed evaluation of the quadratic contribution in the population size $N$.  
We do not report the calculation, which is a trivial application of the perturbative 
theory very much similar to the one performed for the linear approximation in $N$, and 
which consists in retaining all the terms of order $N^2(\eta^0)^4/\sigma^4$ out of 
the expansion of the logarithm.

The final expression for the second order contributions to the information in either case 
reads:

\begin{eqnarray}
&&I_2
\simeq-\frac{1}{\ln2}\frac{N^2(\eta^0)^4}{2(4\sigma^2)^2}
\left\{\frac{p-1}{p^2}2\left(2\left(\alpha-A_1\right)^2\lambda_1\right)^2
+\frac{(\lambda_2)^2}{p}
\left[\left(\frac{1}{2^{2m-1}}\right)^4
\sum_{\nu=0}^{m-1}\left[\left(\begin{array}{c}2m\\\nu\end{array}
\right)\right]^4\right]\right\};\nonumber
\label{final_info4}
\end{eqnarray}

\begin{eqnarray}
&&I_2^{corr}
\simeq-\frac{1}{\ln2}\frac{N^2(\eta^0)^4}{2(4\sigma^2)^2}
\left\{\frac{p-1}{p^2}2\left(2\left(\alpha-A_1\right)^2\lambda_1\right)^2
+\left[\frac{p-1}{p}\left(\lambda_1-\lambda_2\right)^2
+\frac{(\lambda_2)^2}{p}\right]\right.\nonumber\\
&&\left.\left[\left(\frac{1}{2^{2m-1}}\right)^4
\sum_{\nu=0}^{m-1}\left[\left(\begin{array}{c}2m\\\nu\end{array}
\right)\right]^4\right]\right\};\nonumber
\label{final_info4_corr}
\end{eqnarray}
where by $I_2^{corr}$ we mean the quadratic contribution in the correlated case 
$\vartheta^0_{i,s}=\vartheta^0_{i}\,\,\forall s,i$.

The same expression as in eq.(\ref{final_info4_corr}) has been obtained in \cite{vale+01b} by 
means of the replica trick. 

\begin{figure}
\hbox{
\mbox{(a)}
\psfig{figure=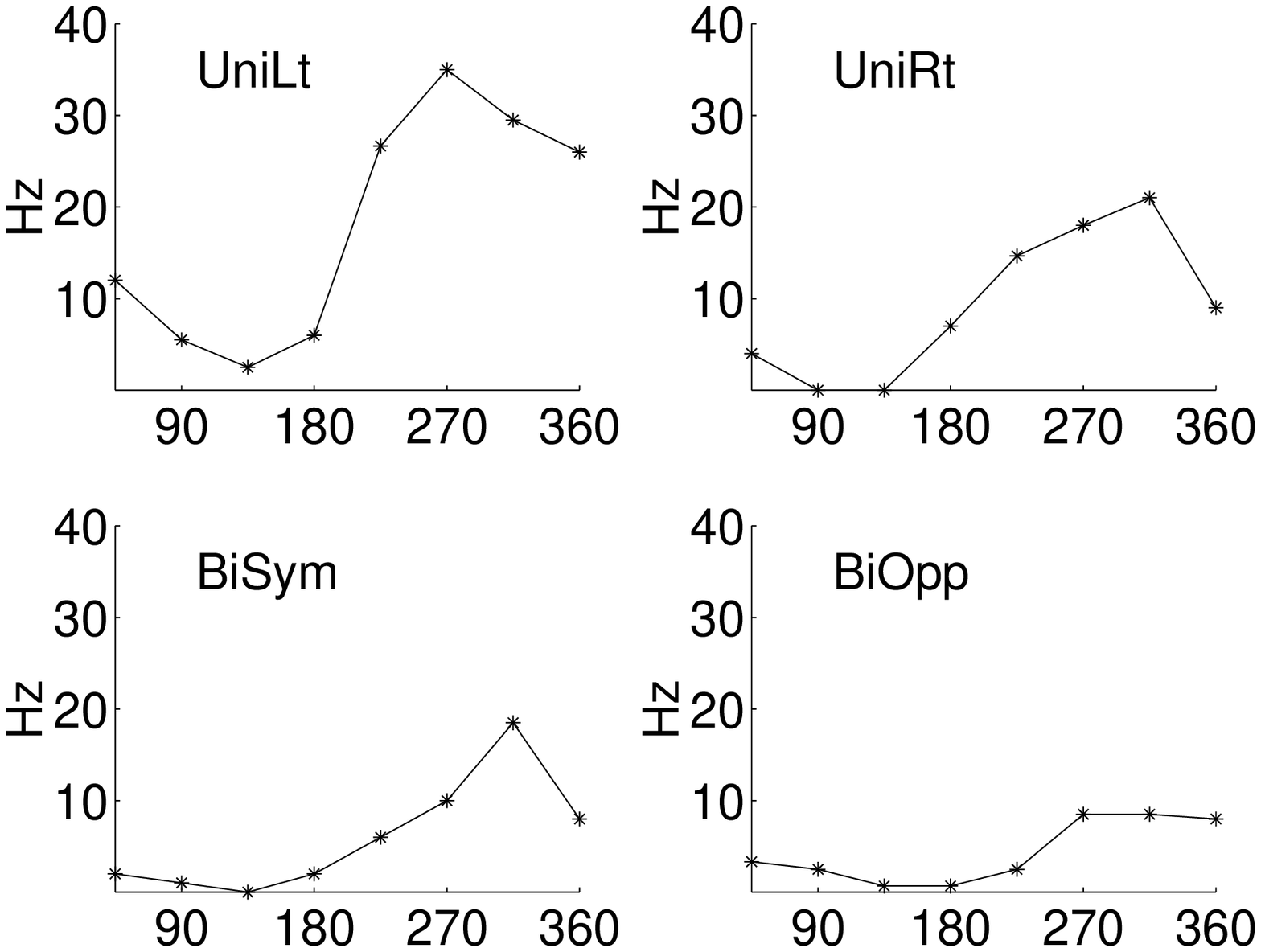,height=6cm,angle=0}
\mbox{(b)}
\psfig{figure=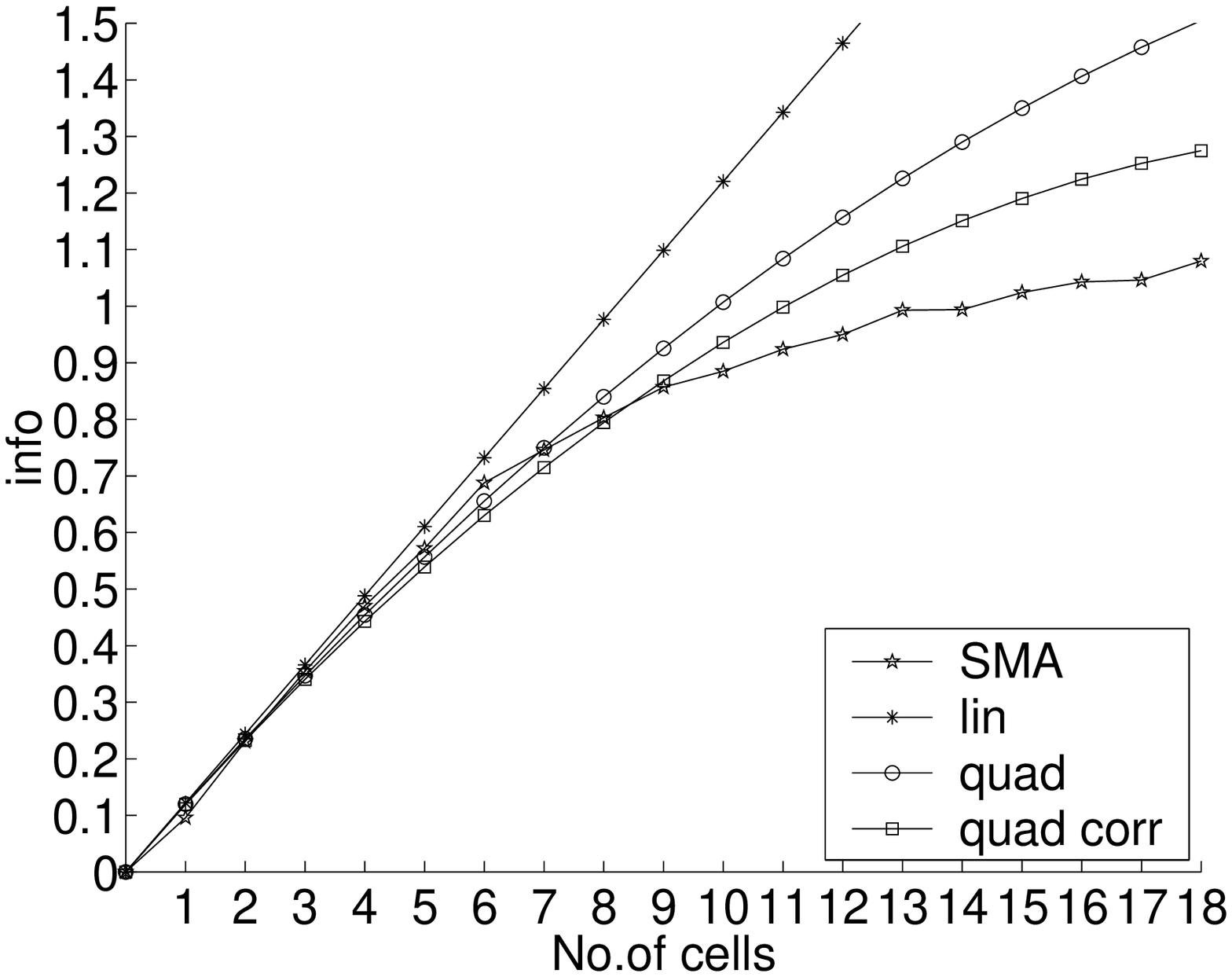,height=6cm,angle=0}}
\caption{
 (a)Directional tuning for a cell recorded in the right supplementary motor area of a monkey performing 4 different types 
of arm movement. UniLt=unimanual left; UniRt=unimanual right; BiSym=bimanual symmetric; BiOpp=bimanual opposite.
reproduced from \protect\cite{vale+01b} 
(b)Comparison between the theoretical curves, eq.(\protect\ref{final_info4}), (\protect\ref{final_info2}), 
(\protect\ref{final_info4_corr}), and the information estimated from a sample of cells 
recorded in the right supplementary motor area \protect\cite{vale}; m=1;p=2; 
the distribution $\varrho(\varepsilon)$ in eqs.(\protect\ref{ro_eps2}),
is just equal to $1/3$ for each of the three allowed 
$\varepsilon$ of $0$,$1/2$,$1$; $\left(\eta^0/2\sigma\right)^2=0.64$.}
\label{fig1}
\end{figure}

Fig.\ref{fig1}, on the left, 
shows an example neuron recorded in the SMA area, whose preferred 
direction does not modulate with the discrete dimension 
(reproduced from \cite{vale+01b}).
Restrictedly to this data set such neurons were statistically dominant, 
even though the significance of such observation should be quantified by means of the analysis of other samples of cells.

On the right we show the theoretical curves in the quadratic approximation corresponding 
to the best linear fit, both for the correlated and uncorrelated case. The curves are compared 
to the information as estimated from a population of SMA cells, showing that the introduction
 of correlations in the preferred directions 
improves the fit. Even far from proving that this precise type of signal 
correlation is the actual mechanism used by SMA cells, this result suggests that 
real cells transmit information firing in a correlated way.

\subsection{Coding of mixed continuous and discrete stimuli 
with a thresholded-gaussian model}
Till now we have examined the information carried about stimuli characterized by 
discrete or mixed continuous and discrete dimensions, assuming that the firing 
of different cells 
is independent across cells and gaussian distributed.
Yet, as already remarked, both assumptions provide a rough approximation of the firing distribution of real 
neurons. The most unjustified one seems to be the gaussian assumption, since it implies that 
also negative rates have a non zero probability of occurrance; a priori the information rise 
might result more or less seriously distorted.

This question has been investigated in \cite{vale+01b}, where a more realistic model 
has been 
proposed, truncating the gaussian distribution and adding a delta peak in zero.

Let us consider once again a population of independent units firing to 
mixed continuous and discrete stimuli $\vartheta\otimes s$, where the single neuron distribution 
is written as follows:

\begin{equation}
P(\eta_i|\vartheta,s)=\frac{1}{\sqrt{2\pi\sigma^2}}
exp-\left[\left(\eta_i-\tilde{\eta}_i(\vartheta,s)
\right)^2/2\sigma^2\right]\Theta(\eta_i)+2(1-\erf(\tilde{\eta}_i(\vartheta,s)/\sigma))
\delta(\eta_i)\Theta(-\eta_i)
\label{dist_corr}
\end{equation}
$\Theta(x)$ is the Heaviside step function and $\tilde{\eta}_i(\vartheta,s)$ has already 
been defined in eq.(\ref{tuning_tot}). $\erf(x)$ is the error function:

\begin{equation}
\erf(x)=\frac{1}{\sqrt{2\pi}}\int_{-\infty}^{x} dt\,\, e^{-t^2/2}.
\label{erf}
\end{equation}

In \cite{vale+01b} the mutual information $I(\{\eta_i\},\vartheta\otimes s)$ has been evaluated 
by means of the replica trick, in the limit of large noise $\sigma$. The interest in this 
limit arises since the larger is $\sigma$, the larger is the gaussian weight assigned as a whole 
to negative 
rates; a consequence might be a larger distortion in the information values.

We show here how the same results can be obtained without the use of the replica trick.

As usual, the information can be expressed as the difference between the equivocation and 
the output entropy, analogously to eqs.(\ref{equiv}),(\ref{outent}) and considering 
the replacements (\ref{rep}).
In \cite{vale+01b} it is shown that the equivocation can be calculated quite easily as a sum of 
single neuron terms; assuming as usual that quenched disorder in uncorrelated and identically 
distributed across stimuli and neurons, according to eqs.(\ref{ro_eps2}),(\ref{ro_0}), one obtains:

\begin{eqnarray}
&&\left\langle H(\{\eta_i\}|\vartheta,s)\right\rangle_{\vartheta,s}=
\frac{N}{2\ln2}\left\{\left(1+\ln(2\pi\sigma^2)\right)
\left\langle\erf(\tilde{\eta}(\vartheta,s)/\sigma)\right\rangle_{\varepsilon,\vartheta^0}
-\left\langle\frac{\tilde{\eta}(\vartheta,s)}{\sqrt{2\pi}\sigma}
e^{-\left[\tilde{\eta}(\vartheta,s)\right]^2/2\sigma^2}
\right\rangle_{\varepsilon,\vartheta^0}\right.\nonumber\\
&&\left.+2\left\langle\left[1-\erf(\tilde{\eta}(\vartheta,s)/\sigma)\right]
\right\rangle_{\varepsilon,\vartheta^0}\ln\frac{\epsilon}{2}-
2\left\langle\left[1-\erf(\tilde{\eta}(\vartheta,s)/\sigma)\right]
\ln\left[1-\erf(\tilde{\eta}(\vartheta,s)/\sigma)\right]
\right\rangle_{\varepsilon,\vartheta^0}\right\}.
\label{eq_need_appr}
\end{eqnarray}
where we have used the representation of the delta function:

\begin{equation}
\int_{-\infty}^{+\infty} dx \delta(x) F(x)=\lim_{\epsilon\rightarrow 0}\int_{-\epsilon/2}^{\epsilon/2} dx \frac{1}{\epsilon} F(x).
\label{delta}
\end{equation}
The equivocation diverges when $\epsilon\rightarrow 0$, but, 
as we will show later on, this divergence 
is canceled exactly by a corresponding term in the entropy of the responses, yielding 
a finite result for the mutual information.

The average across quenched disorder can be performed in the limit of large 
$\sigma$ and expanding the error functions in eq.(\ref{eq_need_appr}) in powers of 
$1/\sigma$.
First we evaluate the entropy of the responses, since we will show that there is a 
partial cancellation of terms.

Considering eq.(\ref{outent}) for the entropy 
of the responses and the replacements
(\ref{rep}), it is easy to show that in the case of the distribution (\ref{dist_corr})  
 one obtains:
\begin{eqnarray}
&&H(\{\eta_i\})=-\left\langle\sum_{s}p(s)\int d\vartheta p(\vartheta)\right.\nonumber\\
&&\left. 
\int \prod_i d\eta_i 
\prod_i \left[\frac{1}{\sqrt{2\pi\sigma^2}}
exp-\left[\left(\eta_i-\tilde{\eta}_i(\vartheta,s)
\right)^2/2\sigma^2\right]\Theta(\eta_i)+2(1-\erf(\tilde{\eta}_i(\vartheta,s)/\sigma))
\delta(\eta_i)\Theta(-\eta_i)\right]\right.\nonumber\\
&&\left.\log_2\left[\sum_{s^\prime}\int d\vartheta^\prime p(\vartheta^\prime,s^\prime)\right.\right.\nonumber\\
&&\left.\left.\prod_i
\left[\frac{1}{\sqrt{2\pi\sigma^2}}
exp-\left[\left(\eta_i-\tilde{\eta}_i(\vartheta^\prime,s^\prime)
\right)^2/2\sigma^2\right]\Theta(\eta_i)+2(1-\erf(\tilde{\eta}_i(\vartheta^\prime,s^\prime)/\sigma))
\delta(\eta_i)\Theta(-\eta_i)
\right]\right]\right\rangle_{\varepsilon,\vartheta^0}.
\label{outent3}
\end{eqnarray}

Developing the products on the neuron index $i$ and taking into account the symmetry 
in the distribution of different units, eq.(\ref{outent3}) can be rewritten as follows:

\begin{eqnarray}
&&H(\{\eta_i\})=-\left\langle\sum_{s}p(s)\int d\vartheta p(\vartheta)\sum_{k=0}
^N 
\left(\begin{array}{c}
N\\k\end{array}
\right)
\int_0^\infty \prod_{i=1}
^k d\eta_i \int_{-\infty}^0 \prod_{i=k+1}
^N d\eta_i\right.\nonumber\\
&&\left.
\prod_{i=1}
^k \frac{1}{\sqrt{2\pi\sigma^2}}
exp-\left[\left(\eta_i-\tilde{\eta}_i(\vartheta,s)
\right)^2/2\sigma^2\right]2^{N-k}
\prod_{i=k+1}
^N 
(1-\erf(\tilde{\eta}_i(\vartheta,s)/\sigma)) \delta(\eta_i)
\right.\nonumber\\
&&\left.
\log_2\left[\sum_{s^\prime}\int d\vartheta^\prime p(\vartheta^\prime,s^\prime)\right.\right.\nonumber\\
&&\left.\left.
\left(\prod_{i=1}
^k \frac{1}{\sqrt{2\pi\sigma^2}}
exp-\left[\left(\eta_i-\tilde{\eta}_i(\vartheta^\prime,s^\prime)
\right)^2/2\sigma^2\right]2^{N-k}
\prod_{i=k+1}
^N 
(1-\erf(\tilde{\eta}_i(\vartheta^\prime,s^\prime)/\sigma)) \delta(\eta_i)
\right)\right]\right\rangle_{\varepsilon,\vartheta^0}.
\label{outent4}
\end{eqnarray}
where we have used the following conventions:
\be
\prod_{i=1}
^k x_i=1\,\,\,\,\,\mbox{if}\,\,\,
 k=0;
\ee
\be
\prod_{i=k+1}
^N x_i=1\,\,\,\,\,\mbox{if}\,\,\,
 k=N;
\ee

Extracting from the logarithm all the factors which do not depend on $\vartheta,s$
and integrating them on $\{\eta_i\}$  it is easy to show that the entropy of the responses can be 
expressed as follows:

\begin{eqnarray}
&&H(\{\eta_i\})=\sum_{s}\int d\vartheta p(\vartheta,s)\nonumber\\
&&\left[\sum_{k=1}
^N 
\left(\begin{array}{c}N\\k\end{array}
\right)
\frac{k}{2\ln 2}\left(1+\ln 2\pi\sigma^2\right)\left\langle 
\erf(\tilde{\eta}(\vartheta,s)/\sigma)\right\rangle^k_{\varepsilon,\vartheta^0}
\left\langle 
1-\erf(\tilde{\eta}(\vartheta,s)/\sigma)\right\rangle^{N-k}_{\varepsilon,\vartheta^0}
\right.\nonumber\\
&&\left.+\sum_{k=0}
^{N-1} 
\left(\begin{array}{c}
N\\k\end{array}
\right)
\frac{N-k}{\ln 2}\ln (\epsilon/2)\left\langle 
\erf(\tilde{\eta}(\vartheta,s)/\sigma)\right\rangle^k_{\varepsilon,\vartheta^0}
\left\langle 
1-\erf(\tilde{\eta}(\vartheta,s)/\sigma)\right\rangle^{N-k}_{\varepsilon,\vartheta^0}\right.\nonumber\\
&&\left.
+\sum_{k=1}
^N 
\left(\begin{array}{c}
 N\\k\end{array}
\right)
\frac{k}{2\ln 2} \left\langle \tilde{\eta}^2(\vartheta,s)/\sigma^2
\erf(\tilde{\eta}(\vartheta,s)/\sigma)\right\rangle_{\varepsilon,\vartheta^0}
\left\langle 
\erf(\tilde{\eta}(\vartheta,s)/\sigma)\right\rangle^{k-1}_{\varepsilon,\vartheta^0}
\left\langle 
1-\erf(\tilde{\eta}(\vartheta,s)/\sigma)\right\rangle^{N-k}_{\varepsilon,\vartheta^0}\right.\nonumber\\
&&\left.+\sum_{k=1}
^N 
\left(\begin{array}{c}
 N\\k\end{array}
\right)
\frac{k}{2\ln 2} \left\langle \tilde{\eta}(\vartheta,s)/\sqrt{2\pi}\sigma
e^{-\tilde{\eta}^2(\vartheta,s)/2\sigma^2}\right\rangle_{\varepsilon,\vartheta^0}
\left\langle 
\erf(\tilde{\eta}(\vartheta,s)/\sigma)\right\rangle^{k-1}_{\varepsilon,\vartheta^0}
\left\langle 
1-\erf(\tilde{\eta}(\vartheta,s)/\sigma)\right\rangle^{N-k}_{\varepsilon,\vartheta^0}
\right]\nonumber\\
&&-\left\langle\sum_{s}\int d\vartheta p(\vartheta,s)\right.\nonumber\\
&&\left.\sum_{k=0}
^N 
\left(\begin{array}{c}
 N\\k\end{array}
\right)
\int_0^\infty \prod_{i=1}
^k d\eta_i \prod_{i=1}
^k \frac{1}{\sqrt{2\pi\sigma^2}}
exp-\left[\left(\eta_i-\tilde{\eta}_i(\vartheta,s)
\right)^2/2\sigma^2\right]
\prod_{i=k+1}
^N 
(1-\erf(\tilde{\eta}_i(\vartheta,s)/\sigma)) 
\right.\nonumber\\
&&\left.
\log_2\left[\sum_{s^\prime}\int d\vartheta^\prime p(\vartheta^\prime,s^\prime)
\left(\prod_{i=1}
^k 
exp\left[2\eta_i\tilde{\eta}_i(\vartheta^\prime,s^\prime)
-\tilde{\eta}_i^2(\vartheta^\prime,s^\prime)/2\sigma^2\right]
\prod_{i=k+1}
^N 
(1-\erf(\tilde{\eta}_i(\vartheta^\prime,s^\prime)/\sigma)) 
\right)\right]\right\rangle_{\varepsilon,\vartheta^0}
\label{outent5}
\end{eqnarray}
where we have used the equality (\ref{delta}) and we have assumed that 
quenched disorder is uncorrelated 
across units and stimuli.

Subtracting the equivocation, eq.(\ref{eq_need_appr}) from the entropy of the responses, 
eq.(\ref{outent5}), it is easy to see that after summation on $k$, the first two terms in 
eq.(\ref{outent5}) simplify with analogous terms in eq.(\ref{eq_need_appr}), so that the 
logarithmic divergence for $\epsilon\rightarrow 0$ cancel out.
 
Finally the mutual information can be rewritten as follows:

\ba
&&I(\{\eta_i\},\vartheta\otimes s)=\sum_s\int d\vartheta p(\vartheta,s)
\left[\frac{N}{\ln 2}\left\langle\left[1-\erf(\tilde{\eta}(\vartheta,s)/\sigma)\right]
\ln\left[1-\erf(\tilde{\eta}(\vartheta,s)/\sigma)\right]
\right\rangle_{\varepsilon,\vartheta^0}\right.\nonumber\\
&&\left.+\frac{N}{\ln 2}\left\langle\frac{\tilde{\eta}(\vartheta,s)}{\sqrt{2\pi}\sigma}
e^{-\left[\tilde{\eta}(\vartheta,s)\right]^2/2\sigma^2}
\right\rangle_{\varepsilon,\vartheta^0}+\frac{N}{2\ln 2}
\left\langle\frac{\tilde{\eta}^2(\vartheta,s)}{\sigma^2}
\erf(\tilde{\eta}(\vartheta,s)/\sigma)
\right\rangle_{\varepsilon,\vartheta^0} 
\right]\nonumber\\
&&-\left\langle\sum_{s}\int d\vartheta p(\vartheta,s)\right.\nonumber\\
&&\left.\sum_{k=0}
^N 
\left(\begin{array}{c}
 N\\k\end{array}
\right)
\int_0^\infty \prod_{i=1}
^k d\eta_i \prod_{i=1}
^k \frac{1}{\sqrt{2\pi\sigma^2}}
exp-\left[\left(\eta_i-\tilde{\eta}_i(\vartheta,s)
\right)^2/2\sigma^2\right]
\prod_{i=k+1}
^N 
(1-\erf(\tilde{\eta}_i(\vartheta,s)/\sigma)) 
\right.\nonumber\\
&&\left.
\log_2\left[\sum_{s^\prime}p(s^\prime)\int d\vartheta^\prime p(\vartheta^\prime)
\left(\prod_{i=1}
^k 
exp\left[2\eta_i\tilde{\eta}_i(\vartheta^\prime,s^\prime)
-\tilde{\eta}_i^2(\vartheta^\prime,s^\prime)/2\sigma^2\right]
\prod_{i=k+1}
^N 
(1-\erf(\tilde{\eta}_i(\vartheta^\prime,s^\prime)/\sigma)) 
\right)\right]\right\rangle_{\varepsilon,\vartheta^0}
\label{info_no_appr}
\ea
Eq.(\ref{info_no_appr}) constitutes the final expression for the mutual information in the 
general case.
To proceed with the analytical evaluation one must 
now resort to some approximation. As suggested in \cite{vale+01b} 
in the limit when the noise $\sigma$ is large one can expand the error functions 
in eq.(\ref{info_no_appr}):

\be
\erf(x)\simeq \frac{1}{2}+\frac{1}{\sqrt{2\pi}}x+o(x^2);
\label{erf_exp}
\ee

A Taylor expansion can be performed also for the exponentials under the logarithm.
One has:

\ba
&&\log_2\left[\sum_{s^\prime}\int d\vartheta^\prime p(\vartheta^\prime,s^\prime)
\left(\prod_{i=1}
^k 
exp\left[2\eta_i\tilde{\eta}_i(\vartheta^\prime,s^\prime)
-\tilde{\eta}_i(\vartheta^\prime,s^\prime)^2/2\sigma^2\right]
\prod_{i=k+1}
^N 
(1-\erf(\tilde{\eta}_i(\vartheta^\prime,s^\prime)/\sigma)) 
\right)\right]\nonumber\\
&&\simeq \log_2 \left[\sum_{s^\prime}p(s^\prime)\int d\vartheta^\prime p(\vartheta^\prime)
\left(1+\sum_{i=1}^k \frac{\eta_i\tilde{\eta}_i(\vartheta^\prime,s^\prime)
}{\sigma^2}-\sum_{i=1}^k\frac{\tilde{\eta}^2_i(\vartheta^\prime,s^\prime)
}{2\sigma^2}+\frac{1}{2}\sum_{i,j=1}^k\frac{
\eta_i\eta_j\tilde{\eta}_i(\vartheta^\prime,s^\prime)
\tilde{\eta}_j(\vartheta^\prime,s^\prime)
}{\sigma^4}\right)\right.\nonumber\\
&&\left. \frac{1}{2^{N-k}}\left(1-\sum_{i=k+1}^{N}\sqrt{\frac{2}{\pi}}
\frac{\tilde{\eta}_i(\vartheta^\prime,s^\prime)}{\sigma}+\frac{1}{\pi}\sum_{i\neq j}
\frac{\tilde{\eta}_i(\vartheta^\prime,s^\prime)
\tilde{\eta}_j(\vartheta^\prime,s^\prime)}{\sigma^2}\right)\right]
\label{log_exp_2}
\ea
Developing all the products one can expand the logarithm again in powers of $1/\sigma$, 
very much similarly to what has been shown in eq.(\ref{log_exp}). The result is then 
integrated on $\{\eta_i\}$ and the binomial sums in eq.(\ref{info_no_appr}) 
can be performed.
Details about the evaluation are given in appendix \ref{appB}.

Finally the mutual information can be written as follows:

\ba
&&I(\{\eta_i\},\vartheta\otimes s)\nonumber\\
&&\simeq\frac{1}{\ln 2}\frac{N}{2\sigma^2}
\left(\frac{1}{2}+\frac{1}{\pi}\right)
\sum_s\sum_{s^\prime} p(s) p(s^\prime)\int d\vartheta \int d\vartheta^\prime 
p(\vartheta) p(\vartheta^\prime) \left[\left\langle
\left(\tilde{\eta}(\vartheta,s)\right)^2
\right\rangle_{\varepsilon,\vartheta^0}-
\left\langle\tilde{\eta}(\vartheta,s)\tilde{\eta}(\vartheta^\prime,s^\prime)
\right\rangle_{\varepsilon,\vartheta^0}\right]\nonumber\\
&&=\frac{1}{\ln2}\left(\frac{1}{2}+\frac{1}{\pi}\right)
\frac{N(\eta^0)^2}{4\sigma^2}
\left[\frac{p-1}{p}2\left(\alpha-A_1\right)^2\sigma^2_\varepsilon+2\left(A_2-(A_1)^2\right)
\left\langle\varepsilon^2\right\rangle_\varepsilon\right];
\label{info_large_corr}
\ea
where we have used eqs.(\ref{Lambda1}),(\ref{Lambda2}) and $A_1$,$A_2$ and $\alpha$ have 
been defined in eq.(\ref{A_2}).

This result equals the expression obtained using replicas in \cite{vale+01b}, showing that 
even with a more complicated distribution, other than the simple gaussian model, 
the evaluation can be carried out via a simple Taylor 
expansion of the logarithm, and no significant advantage derives from the use of the replica trick 
in the limit case of large noise.

Comparing eqs.(\ref{info_large_corr}) and (\ref{final_info2}) one can notice that limitedly 
to the case of large noise, the effect 
of thresholding the gaussian distribution with respect to the information 
is merely a renormalization 
of the noise for a factor $1/\sqrt{1/2+1/\pi}$. 
In \cite{vale+01b} it has been shown that this renormalization effect 
holds at higher orders in $1/\sigma^2$.


\section{Population information in the asymptotic regime of large $N$ and small $\sigma$}
We turn now to the analysis of the asymptotic regime in the information curve, for a large 
number of neurons. 
A first attempt to solve this limit in the case of independent gaussian units and discrete stimuli, 
as in eq.(\ref{dist}) has been done in \cite{pap35} by means of the replica trick.
Yet, probably due to some too strong approximation in summing on replicas, the 
final analytical expression was incorrect, according to the authors.
 
The asymptotic behaviour was then studied in \cite{sompo+00} distinctly in the 
case of discrete and continuous 
stimuli, and for a generic distribution of independent units, yet in absence of additional quenched 
disorder.
We try here to go further and study the case where quenched disorder is present, as in 
distribution (\ref{dist}), and mixed continuous and discrete dimensions 
characterize simultaneously the stimulus structure.
We focuse first on the simpler case of purely discrete stimuli. 
In this context we compare an approach which is equivalent in the nature of the approximation 
to the one presented in \cite{pap35}, yet replica free, to the 
approach presented in \cite{sompo+00}. We show in detail under which approximation 
the two approaches provide the same result.

\subsection{Coding of discrete stimuli in a gaussian approximation}

Let us reconsider eq.(\ref{no_appr}). When $\sigma$ becomes very small the probability density $p(\eta_i|s)$ can be approximated by a $\delta$-function:

\be
\frac{e^{-\left(\eta_i-\eta_i^s\right)^2/2\sigma^2}}{\sqrt{2\pi\sigma^2}}
\longrightarrow \delta\left(\eta_i-\eta_i^s\right);
\label{ines_appr}
\ee
This approximation, which corresponds to freezing the quenched disorder represented 
by the variables  $\{\eta_i\}$ under the logarithm, has been used in \cite{pap35} after 
getting rid of the logarithm itself by means of the replica trick.
Under this approximation the integration on $\{\eta_i\}$ can be performed and the 
mutual information can be rewritten as follows:

\ba
&&I(\{\eta_i\},s)\simeq -\frac{1}{\ln 2}\left\langle \sum_{s}p(s)
\ln\left[\sum_{s^\prime} p(s^\prime) 
\prod_i e^{-(\eta_i^s-\eta_i^{s^\prime})^2/2\sigma^2}\right]\right\rangle_\eta\nonumber\\
&&=\log_2 p-\frac{1}{\ln 2}\left\langle \sum_{s}p(s)
\ln\left[1+\sum_{s^\prime\neq s}  
\prod_i e^{-(\eta_i^s-\eta_i^{s^\prime})^2/2\sigma^2}\right]\right\rangle_\eta;
\label{small_sigma}
\ea
where upper bound $\log_2 p$ derives from the term with $s=s^\prime$ in the sum on 
$s^\prime$ and we have used $p(s)=const.=1/p$.

Since $\sigma$ is small one can expand the logarithm:
\ba
&&\ln\left[1+\sum_{s^\prime\neq s}  
\prod_i e^{-(\eta_i^s-\eta_i^{s^\prime})^2/2\sigma^2}\right]\nonumber\\
&&
\simeq \sum_{s^\prime\neq s}\prod_i e^{-(\eta_i^s-\eta_i^{s^\prime})^2/2\sigma^2}+
\frac{1}{2}\sum_{s^\prime\neq s}\sum_{s^{\prime\prime}\neq s}\prod_{ij}
e^{-(\eta_i^s-\eta_i^{s^\prime})^2/2\sigma^2}
e^{-(\eta_j^s-\eta_j^{s^{\prime\prime}})^2/2\sigma^2}+...
\label{small_sigma_exp}
\ea

In the appendix we show that inserting this expansion in eq.(\ref{small_sigma}) one can perform the quenched averages
and derive an explicit expression for the mutual information. The final result reads:

\be
I(\{\eta_i\},s)\simeq \log_2 p\left[1-\frac{p-1}{\log_2p\ln 2}
\left(S_1\left(\sqrt{2\pi}\sigma I_1\right)^N-\left(p-2\right)S_2
\left(2\pi\sigma^2I_2\right)^N\right)\right];
\label{final_info}
\ee
where we have considered the leading term order $\sigma$ and the first correction 
of order $\sigma^2$, and one has:
\ba
&&S_1=\sum_{k=1}^{\infty}\frac{\left(-1\right)^{k+1}}{k^{N/2+1}};\,\,\,\,\,\,\,\,\,\,\,
S_2=\sum_{k=1}^{\infty}\frac{\left(-1\right)^{k+1}}{k^{\frac{N}{2}}};\nonumber\\
&&I_1=\int d\eta \varrho^2(\eta);\,\,\,\,\,\,\,\,\,\,\,I_2=\int d\eta \varrho^3(\eta);
\ea

When the noise goes to zero and the population size is large the information reaches 
the upper bound of $\log_2 p$.

\begin{figure}
\psfig{figure=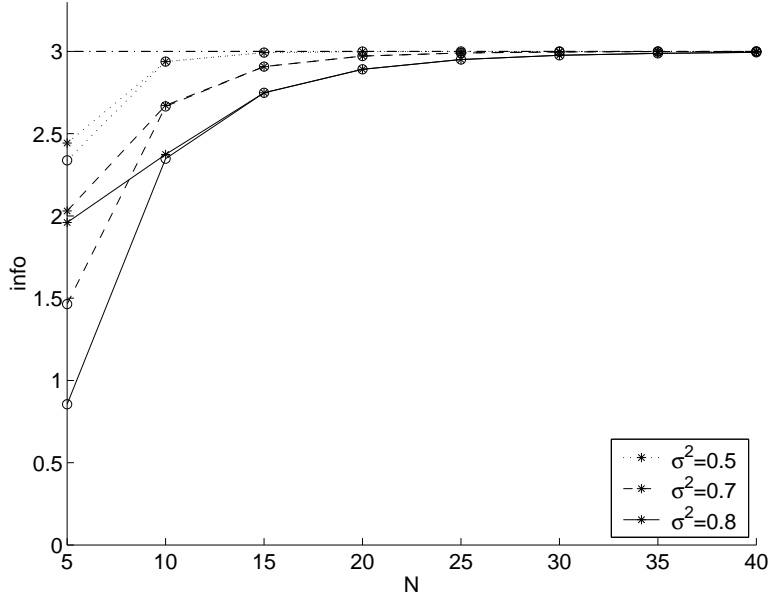,height=8cm,angle=0}
\caption{Mutual information as in eq.(\protect\ref{final_info}), as a function of the population size
$N$. Different curves correspond to different values of the noise $\sigma$; circles 
are for the information with only the leading term of order $\sigma^N$ in eq.(\protect\ref{final_info}), while 
stars are for the full information with also the correction of order $\sigma^{2N}$}
\label{fig3}
\end{figure}

Fig.\ref{fig3} shows the mutual information according to eq.(\ref{final_info})
as a function of the population size and for different values of the noise $\sigma$.
Circles and stars are respectively for the full mutual information with both the 
leading and 
the correction terms in eq.(\ref{final_info}) and with only the leading term of order 
$\sigma^N$. 
As it is evident from the plot, the larger $\sigma$, the slower the approach to the ceiling, and 
the larger the weight acquired by the correction term of order $\sigma^{2N}$.

An alternative replica-free method to study the asymptotic information regime has been proposed in \cite{sompo+00}. 
In principle the method looks 
quite efficient and moreover it can be applied both to continuous and to discrete stimuli.

Yet no additional quenched disorder affected the distributions considered in \cite{sompo+00}.
We try now to apply the method to our particular coding scheme.

Let us reconsider eq.(\ref{info}) for the mutual information. 
With a change of variables it can be rewritten:
\be
I(\{\eta_i\},s)=-\sum_{s} p(s)
\int \prod_{s^{\prime\prime}\neq s} d X_{s^{\prime\prime}}
\left\langle p(\{X_{s^{\prime\prime}}\})\right\rangle_\eta
\log_2\left[\sum_{s^{\prime}} p(s^\prime)\exp\left(X_{s^\prime}\right)\right];
\label{info_x} 
\ee
\be
X_{s^{\prime}}=\frac{1}{N}\ln\left(\frac{p(\{\eta_i\}|s^\prime)}{p(\{\eta_i\}|s)}\right)=
\frac{1}{N}\sum_i(\eta_i^s-\eta_i^{s^\prime})(\eta_i^s+\eta_i^{s^\prime}-2\eta_i)/2\sigma^2;
\ee
\be
p(\{X_{s^{\prime}}\})=\int d \{\eta_i\} p(\{\eta_i\}|s)\prod_{s^{\prime}\neq s} 
\delta\left(X_{s^{\prime}}-X_{s^{\prime}}\left(\{\eta_i\},s,s^\prime\right)\right);
\label{p_X}
\ee
where in deriving the explicit expression for $X_{s^{\prime}}$ we have used the distribution 
(\ref{dist}).
This change of variable allows to move the quenched disorder from inside the logarithm 
to the distribution $p(\{X_{s^{\prime}}\})$.

Using the integral representation for the $\delta$ function and integrating on 
$\{\eta_i\}$ eq.(\ref{p_X}) can be rewritten as follows:
\ba
&&p(\{X_{s^{\prime}}\})=\int \prod_{s^{\prime}\neq s} \frac{d Y_{s^{\prime}}}{2\pi} 
\exp\left(-\imath\,\, Y_{s^{\prime}}\left[X_{s^{\prime}}
+\frac{1}{N}\sum_i (\eta_i^{\prime}-\eta_i^s)^2/2\sigma^2\right]\right)\nonumber\\
&&\prod_{s^{\prime}\neq s}\prod_{s^{\prime\prime}\neq s}  
\exp\left(-\frac{Y_{s^{\prime}}}{N}\frac{Y_{s^{\prime\prime}}}{N}\sum_i 
(\eta_i^{s^{\prime}}-\eta_i^s)(\eta_i^{s^{\prime\prime}}-\eta_i^s)/2\sigma^2\right);
\label{p_X2}
\ea
We notice that when $N$ grows large and $\sigma$ becomes small the argument in the 
second exponential is of order $1/N$ with respect to the first one: 
a relatively flat gaussian is multiplied times a strongly oscillating periodic 
function. Thus we put the second function equal to one obtaining:
\be
p(\{X_{s^{\prime}}\})=\prod_{s^{\prime}\neq s} \delta\left(X_{s^{\prime}}
+\frac{1}{N}\sum_i (\eta_i^{\prime}-\eta_i^s)^2/2\sigma^2\right);
\label{p_X_sompo}
\ee
We notice that:
\ba
&&\frac{1}{N}\sum_i (\eta_i^{\prime}-\eta_i^s)^2/2\sigma^2=
-\int d \{\eta_i\} p(\{\eta_i\}|s)X_{s^{\prime}}\left(\{\eta_i\},s,s^\prime\right)\nonumber\\
&&=-\int d \{\eta_i\} p(\{\eta_i\}|s)
\frac{1}{N}\ln\left(\frac{p(\{\eta_i\}|s^\prime)}{p(\{\eta_i\}|s)}\right)=
-\frac{1}{N}KL(s^\prime ||s);
\ea
where $KL(s^\prime ||s)$ is the Kullback-Leibler divergence between the distributions 
$p(\{\eta_i\}|s^\prime)$ and $p(\{\eta_i\}|s)$.
Thus under this approximation the distribution $p(\{X_{s^{\prime}}\})$ factorizes 
into a product of delta-functions centered on the mean value of 
$X_{s^{\prime}}\,\,\forall s^\prime$.

Inserting eq.(\ref{p_X_sompo}) in the expression for the mutual information, eq.(\ref{info_x}), it is 
easy to show that integrating on $\{X_{s^{\prime}}\}$ one reobtains eq.(\ref{small_sigma}).
Thus, approximating the variables $\{X_{s^{\prime}}\}$ with the mean of the distribution, as in eq.(\ref{p_X_sompo}), 
corresponds to the $\delta$-function approximation, eq.(\ref{ines_appr}).

A more accurate estimate might be obtained calculating the correction given by 
the second exponential in eq.(\ref{p_X2}), that we had previously neglected.
Yet, integration on $\{Y_{s^{\prime}}\}$ would lead to the introduction of matrices which depend 
on the quenched disorder $\{\eta_i^s\}$, which therefore must be averaged out first, in order to derive 
$\left\langle p(\{X_{s^{\prime\prime}}\})\right\rangle_\eta$.
The average might be performed specifying the distribution of the quenched disorder $\varrho(\eta)$, but even in this case, 
integrating out the variables $\{\eta_i^s\}$ introduces a non trivial dependence on the variables $\{Y_{s^{\prime}}\}$ 
which in turn can be integrated out by means of further ansatz. 
Details about the evaluation of the correction and of the results will be published elsewhere \cite{vale+03a}.

\subsection{Coding of mixed discrete and continuous stimuli: gaussian vs thresholded gaussian model}

As we have shown in the previous section, in the large $\sigma$ limit both the technical evaluation and the final 
expression for the mutual information are formally the same whether the stimuli are discrete or continuous. 
On the other hand in the asymptotic regime of small $\sigma$ a qualitative difference exists between the case where the 
stimuli are discrete or continuous, in that in the former case the information is bounded by the entropy of the 
stimulus set, while in the latter the information grows to infinity as the noise goes to zero, or if the number of 
neurons grows to infinity for a finite noise.
Here we calculate the expression of this asymptotic growth, which corresponds to the upper bound in the case of purely 
discrete stimuli. We compare the two cases of gaussian and thresholded gaussian models, in order to assess whether, as
in the large $\sigma$ regime, a plain relationship like a renormalization of the noise links the two expressions. 

Let us reconsider the distribution (\ref{dist2}). 
The expression for the mutual information is given by eq.(\ref{no_appr2}), that we recall:

\ba
&&I(\{\eta_i\},\vartheta\otimes s)=-\left\langle \frac{1}{\ln 2} \sum_{s}p(s)
\int d\vartheta p(\vartheta) \int \prod_i d\eta_i \prod_i 
e^{-\left(\eta_i-\tilde{\eta_i}(\vartheta,s)\right)^2/2\sigma^2}\right.\nonumber\\
&&
\left.\ln\left[\sum_{s^\prime} p(s^\prime) \int d\vartheta^\prime
\prod_i e^{-\left(\left(\tilde{\eta}_i(\vartheta,s)\right)^2+
\left(\tilde{\eta}_i\left(\vartheta^\prime,s^\prime\right)\right)^2
-2\eta_i\tilde{\eta}_i\left(\vartheta^\prime,s^\prime\right)\right)/2\sigma^2}\right]
\right\rangle_{\varepsilon,\vartheta^0};
\label{no_appr2_bis}
\ea
 
In analogy with the approximation (\ref{ines_appr}), in the limit when $\sigma$ becomes very small 
we use:

\be
\frac{e^{-\left(\eta_i-\tilde{\eta}_i(\vartheta,s)\right)^2/2\sigma^2}}{\sqrt{2\pi\sigma^2}}
\longrightarrow \delta\left(\eta_i-\tilde{\eta}_i(\vartheta,s)\right);
\label{ines_appr2}
 \ee
Under this aproximation we obtain in analogy to eq.(\ref{small_sigma}):

\ba
&&I(\{\eta_i\},\vartheta\otimes s)\simeq -\frac{1}{\ln 2}\left\langle \sum_{s}p(s)\int d\vartheta p(\vartheta)
\ln\left[\sum_{s^\prime} p(s^\prime) \int d\vartheta^\prime p(\vartheta^\prime)
\prod_i e^{-[\tilde{\eta}_i(\vartheta^\prime,s^\prime)-\tilde{\eta}_i(\vartheta,s)]^2/2\sigma^2}\right]\right\rangle_{\vartheta^0,\varepsilon};
\label{small_sigma2}
\ea

In the previous section we have seen that in the case of $p$ purely discrete stimuli the upper bound, $\log_2 p$,  was given 
by the term with $s=s^\prime$ under the logarithm.
Therefore in the case of mixed continuous and discrete stimuli we expect the same term, after integration on $\vartheta^\prime$, 
to give the logarithm 
of a coefficient depending on the ratio between $N$ and $\sigma^2$, which grows to infinity when $\sigma\rightarrow 0$.
Let us extract and calculate the term with $s=s^\prime$ out of eq.(\ref{small_sigma2}):

\be
\int d\vartheta^\prime p(\vartheta^\prime) p(s)
\prod_i e^{-[\tilde{\eta}_i(\vartheta^\prime,s)-\tilde{\eta}_i(\vartheta,s)]^2/2\sigma^2}
\ee
It is clear that when $\sigma$ becomes small the major contribution to the integral comes from the 
values $\vartheta^\prime\simeq\vartheta$. 
Therefore we expand the difference $\tilde{\eta}_i(\vartheta^\prime,s)-\tilde{\eta}_i(\vartheta,s)$:

\be
\tilde{\eta}_i(\vartheta^\prime,s)-\tilde{\eta}_i(\vartheta,s)\simeq \frac{\partial \tilde{\eta}_i(\vartheta^\prime,s)}
{\partial \vartheta^\prime}_{|_{\vartheta^\prime=\vartheta}}(\vartheta^\prime-\vartheta)=\frac{\eta^0}{2}
\varepsilon^i_s\sin(\vartheta-\vartheta^0_{i,s})(\vartheta^\prime-\vartheta);
\label{expansion}
\ee
 where we have explicitely used the expressions (\ref{tuning_tot}),(\ref{tuning2}).

In the limit when $\sigma\rightarrow 0$ the resulting gaussian distribution is approximated with a $\delta$-function 
and the integral on $d\vartheta^\prime$ can be performed. 
Extracting this contribution out of the logarithm the information can be rewritten as follows:

\ba
&&I(\{\eta_i\},\vartheta\otimes s)\simeq \log_2 \left[\frac{p\sqrt{\pi}}{\sqrt{2}}\frac{\sqrt{N}\eta^0}{\sigma}\right]+
\sum_s \int d\vartheta p(s,\vartheta)
\left\langle \frac{1}{2}\log_2\left[\frac{1}{N}\sum_i (\varepsilon^i_s)^2\sin^2(\vartheta-\vartheta^0_{i,s})\right]
\right\rangle_{\vartheta^0,\varepsilon}\nonumber\\
&&-\frac{1}{\ln 2}\left\langle \sum_{s}p(s)\int d\vartheta p(\vartheta)\right.\nonumber\\
&&\left.
\ln\left[1+\frac{\eta^0}{\sqrt{8\pi\sigma^2}}\sqrt{\sum_{i=1}^N (\varepsilon^i_s)^2\sin^2(\vartheta-\vartheta^0_{i,s})}
\sum_{s^\prime\neq s} \int d\vartheta^\prime
\prod_i e^{-[\tilde{\eta}_i(\vartheta^\prime,s^\prime)-\tilde{\eta}_i(\vartheta,s)]^2/2\sigma^2}\right]\right\rangle_\eta;
\label{small_sigma3}
\ea
The quenched average in the second term can be performed in the thermodynamic limit letting the average pass  
the logarithm in a mean field approximation.
The third term behaves like $\log_2 [1+\Delta]$, where $\Delta$ 
vanishes like $\sqrt{N}/\sigma\exp(-N/\sigma^2)$ when $\sigma\rightarrow 0$.
The information in the leading term can be finally expressed in the following form:

\be
I(\{\eta_i\},\vartheta\otimes s)\simeq 
\log_2 \left(\frac{p\sqrt{\pi}}{\sqrt{2}}\frac{\sqrt{N}\eta^0}{\sigma}\sqrt{\frac{\langle \varepsilon^2\rangle_\varepsilon}{2}}\right)
\label{info_asympt_gauss}
\ee

When the number of neurons grows to infinity and/or the noise tends to zero the information grows logarithmically to infinity.
Notice that the case of purely continuous stimuli can be retrieved putting $p=1$: not surprisingly the continuous dimension 
plays a major role in determining the asymptotic growth of the information to infinity, with a relatively mild modulation according to 
the number $p$ of discrete correlates.

We turn now to the case of the thresholded-gaussian model, eq.(\ref{dist_corr}).
As we have shown previously, the mutual information can be expressed as in eq.(\ref{info_no_appr}).
In the limit of small $\sigma$ we apply the approximation (\ref{ines_appr2}). After a rearrangement of the terms 
we obtain:

\ba
&&I(\{\eta_i\},\vartheta\otimes s)\simeq\sum_s\int d\vartheta p(\vartheta,s)
\left[\frac{N}{\ln 2}\left\langle\left[1-\erf(\tilde{\eta}(\vartheta,s)/\sigma)\right]
\ln\left[1-\erf(\tilde{\eta}(\vartheta,s)/\sigma)\right]
\right\rangle_{\varepsilon,\vartheta^0}\right.\nonumber\\
&&\left.+\frac{N}{\ln 2}\left\langle\frac{\tilde{\eta}(\vartheta,s)}{\sqrt{2\pi}\sigma}
e^{-\left[\tilde{\eta}(\vartheta,s)\right]^2/2\sigma^2}
\right\rangle_{\varepsilon,\vartheta^0}+\frac{N}{2\ln 2}
\left\langle\frac{\tilde{\eta}^2(\vartheta,s)}{\sigma^2}
\erf(\tilde{\eta}(\vartheta,s)/\sigma)
\right\rangle_{\varepsilon,\vartheta^0} 
\right]\nonumber\\
&&-\left\langle\sum_{s}\int d\vartheta p(\vartheta,s)
\sum_{k=0}
^N 
\left(\begin{array}{c}
 N\\k\end{array}
\right)
\prod_{i=k+1}
^N 
(1-\erf(\tilde{\eta}_i(\vartheta,s)/\sigma)) 
\left\{\sum_{i=1}^k \frac{k}{\ln 2}
\frac{\tilde{\eta}_i^2(\vartheta,s)}{\sigma^2}\right.\right.\nonumber\\
&&\left.\left.+\log_2\left[\sum_{s^\prime}p(s^\prime)\int d\vartheta^\prime p(\vartheta^\prime)
\left(\prod_{i=1}
^k 
e^{-\left[(\tilde{\eta}_i(\vartheta^\prime,s^\prime)
-\tilde{\eta}_i(\vartheta,s))^2/2\sigma^2\right]}
\prod_{i=k+1}
^N 
(1-\erf(\tilde{\eta}_i(\vartheta^\prime,s^\prime)/\sigma)) 
\right)\right]\right\}\right\rangle_{\varepsilon,\vartheta^0}
\label{info_first_appr}
\ea
Out of this expression we aim at keeping the leading terms, which diverge when $\sigma$ goes to zero and determine the 
asymptotic behaviour. In first approximation we will neglect terms going to zero with $\sigma$, which would play a role 
in the first correction to the asymptotic value.
As already done in the case of the of the gaussian model we consider the term with $s^\prime=s$ in the discrete 
sum under the logarithm.
Looking at the function to be integrated on $\vartheta^\prime$
it is easy to see that the product of the $k$ exponentials is maximal for 
values of $\vartheta^\prime$ close to $\vartheta$, while each one of the other $N-k$ 
factors containing error functions is maximal for cell and stimulus specific values of $\vartheta^\prime$, namely 
the ones corresponding to the smallest values of $\tilde{\eta}_i(\vartheta^\prime,s)$.
Thus the main contribution for the integral comes from the values of $\vartheta^\prime$ close to $\vartheta$ and we can repeat the procedure 
as in eqs.(\ref{expansion}),(\ref{small_sigma3}), obtaining:

\ba
&&I(\{\eta_i\},\vartheta\otimes s)\simeq\sum_s\int d\vartheta p(\vartheta,s)
\left[\frac{N}{\ln 2}\left\langle\left[1-\erf(\tilde{\eta}(\vartheta,s)/\sigma)\right]
\ln\left[1-\erf(\tilde{\eta}(\vartheta,s)/\sigma)\right]
\right\rangle_{\varepsilon,\vartheta^0}\right.\nonumber\\
&&\left.+\frac{N}{\ln 2}\left\langle\frac{\tilde{\eta}(\vartheta,s)}{\sqrt{2\pi}\sigma}
e^{-\left[\tilde{\eta}(\vartheta,s)\right]^2/2\sigma^2}
\right\rangle_{\varepsilon,\vartheta^0}+\frac{N}{2\ln 2}
\left\langle\frac{\tilde{\eta}^2(\vartheta,s)}{\sigma^2}
\erf(\tilde{\eta}(\vartheta,s)/\sigma)
\right\rangle_{\varepsilon,\vartheta^0} 
\right]\nonumber\\
&&-\left\langle\sum_{s}\int d\vartheta p(\vartheta,s)
\sum_{k=0}
^N 
\left(\begin{array}{c}
 N\\k\end{array}
\right)
\prod_{i=k+1}
^N 
(1-\erf(\tilde{\eta}_i(\vartheta,s)/\sigma)) 
\left\{\sum_{i=1}^k \frac{k}{\ln 2}
\frac{\tilde{\eta}_i^2(\vartheta,s)}{\sigma^2}\right.\right.\nonumber\\
&&\left.\left.-\log_2 \left[\frac{p\sqrt{\pi}}{\sqrt{2}}\frac{\sqrt{k}\eta^0}{\sigma}\right]
-\frac{1}{2}\log_2\left[\frac{1}{k}\sum_{i=1}^k (\varepsilon^i_s)^2\sin^2(\vartheta-\vartheta^0_{i,s})\right]
+\sum_{i=k+1}^N \log_2(1-\erf(\tilde{\eta}_i(\vartheta,s)/\sigma))\right.\right.\nonumber\\
&&\left.\left.+\log_2\left[1+\frac{\eta^0}{\sqrt{8\pi\sigma^2}}\sqrt{\sum_{i=1}^k 
(\varepsilon^i_s)^2\sin^2(\vartheta-\vartheta^0_{i,s})}\,\,\mbox{X}\right.\right.\right.\nonumber\\
&&\left.\left.\left.\mbox{X}\,\,\sum_{s^\prime\neq s}\int d\vartheta^\prime 
\left(\prod_{i=1}
^k 
e^{-\left[(\tilde{\eta}_i(\vartheta^\prime,s^\prime)
-\tilde{\eta}_i(\vartheta,s))^2/2\sigma^2\right]}
\prod_{i=k+1}
^N 
(1-\erf(\tilde{\eta}_i(\vartheta^\prime,s^\prime)/\sigma)) 
\right)\right]\right\}\right\rangle_{\varepsilon,\vartheta^0}
\label{info_second_appr}
\ea
Performing the sums on $k$ and taking the limit of small $\sigma$ it is easy to show that all terms simplify or vanish 
with higher orders in $\sigma$,
except for the second and third term in the sum on $k$. The third term can be evaluated in a mean field approximation 
passing the quenched average under the logarithm: since $N$ is very large $k$ is also large for most values.
The leading asymptotic term finally reads:

\ba
&&I(\{\eta_i\},\vartheta\otimes s)\simeq 
\log_2 \left(\frac{p\sqrt{\pi}}{\sqrt{2}}\frac{\sqrt{N}\eta^0}{\sigma}
\sqrt{\frac{\langle \varepsilon^2\rangle_\varepsilon}{2}}\right)\nonumber\\
&&+\frac{1}{2}\sum_{s}\int d\vartheta p(\vartheta,s)
\sum_{k=1}
^{N-1} 
\left(\begin{array}{c}
 N\\k\end{array}
\right)
\left\langle 1-\erf(\tilde{\eta}_i(\vartheta,s)/\sigma)\right\rangle^{N-k}_{\varepsilon,\vartheta^0}
\log_2 \left(\frac{p\sqrt{\pi}}{\sqrt{2}}\frac{\sqrt{k}\eta^0}{\sigma}
\sqrt{\frac{\langle \varepsilon^2\rangle_\varepsilon}{2}}\right);
\label{info_asympt_th}
\ea
Since the first term is exactly the information for the gaussian model, eq.(\ref{info_asympt_gauss}), we see that already 
at the leading term, the rise to infinity with $N$ is slightly higher for the thresholded gaussian model.
It is quite difficult to extract analytically the exact dependence on $N$ and 
$\sigma$ from eq.(\ref{info_asympt_th}).
We have evaluated the sum on $k$ numerically by means of a MATLAB code.

\begin{figure}
\hbox{
\mbox{(a)}
\psfig{figure=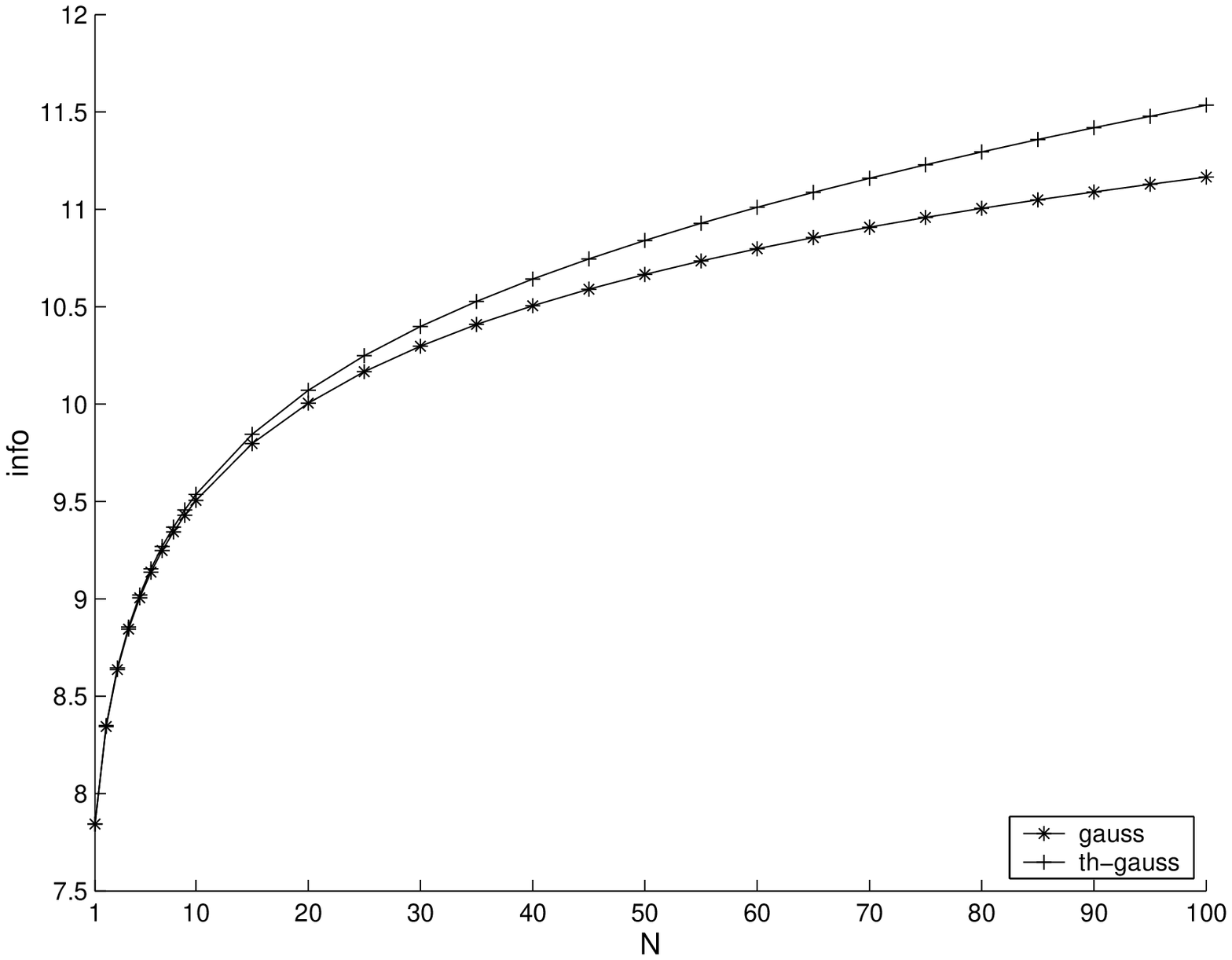,height=6cm,angle=0}
\mbox{(b)}
\psfig{figure=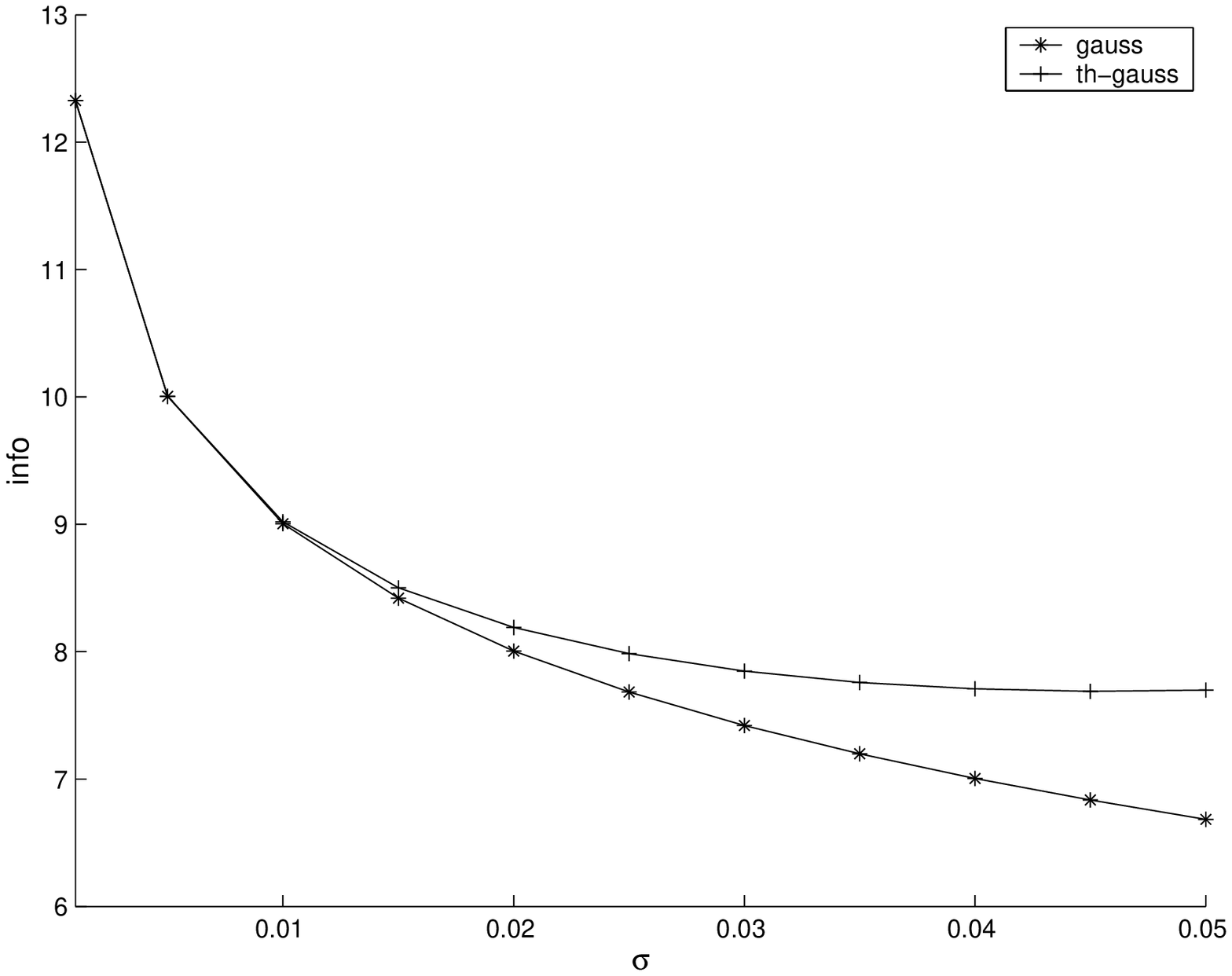,height=6cm,angle=0}}
\caption{Information for the gaussian vs thresholded gaussian model with mixed continuous and discrete stimuli, 
according to eqs.(\protect\ref{info_asympt_gauss}),(\protect\ref{info_asympt_th}). $\eta_0=1$; $p=4$;$\alpha=0.2$;
$\varepsilon^i_s$ can take values $3/10$,$6/10$ and $9/10$ with equal probability.
 (a)Asymptotic behaviour as a function of the population size $N$ for $\sigma=0.01$.
(b)As a function of the noise $\sigma$ for a population size of $N=5$ cells.}
\label{fig4}
\end{figure}

Fig.(\ref{fig4})(a) shows the mutual information for both models as a function of the population size. 
It must be said that while eq.(\ref{info_asympt_gauss}) is valid for generic values of the parameters, provided 
the noise is small, more restrictive assumptions underlie the derivation of eq.(\ref{info_asympt_th}).
In particular one has to exclude the values for which the tuning curve $\tilde{\eta}_i(\vartheta,s)$
is identically zero (namely $\alpha=0$ and $\varepsilon_s^i=1$ with a finite weight in 
eq.(\ref{tuning_tot})); the reason is that for very small values of the noise the weight of the $\delta$
peak in the distribution (\ref{dist_corr}) becomes proportional to $\delta(\tilde{\eta}_i(\vartheta,s))$, and 
several terms in eq.(\ref{info_second_appr}) cannot be neglected any more.

Moreover, for any fixed value of $N$ there is an upper bound on the value of the noise, beyond which the approximation 
(\ref{info_asympt_gauss}) must be integrated by the neglected terms. 
This can be seen by direct analytical evaluation of each term. 

As an example we show in fig.(\ref{fig4})(b) this effect for a population of $5$ cells: 
for values of $\sigma$ close to $0.05$ the decrease in the information 
for the thresholded gaussian model starts slowing down, and for higher values of the noise the information 
would even increase.

\section{Discussion}
We have presented a detailed analysis of the mutual information carried by one population of independent units 
about a set of stimuli, examining both the case where the stimuli are purely discrete and the case where they are characterized 
by an additional continuous angular dimension. 
In fact, even though in real experiments performed on trained animals the stimuli always vary in a discrete set, since even continuous 
dimensions are sampled on a finite number of points, in nature the real brain must learn how to discriminate 
highly dimensional stimuli 
or behavioural correlates, whose dimensions may equally be continuous or discrete. 
For our specific model, we have been inspired by data recorded in the motor cortex of monkeys performing arm movements 
which might be parameterized according to a (continuous) direction and to a (discrete) "type" \cite{Don+98,vale}, 
We have focused on two possible limits, namely the limit of finite population size and large noise, which corresponds to the initial 
information rise, and the asymptotic regime of large numbers of neurons and small noise.
The limit of large noise has been recently studied in \cite{vale+01b} by means of the replica trick.
Here we have shown that regardless of the structure of the stimulus whether continuous or discrete, the same results can be 
obtained without resorting to the replica trick, by a mere expansion of the logarithm.
Moreover we have shown that correlations introduced in the preferred orientations of each neuron, across different values of 
the discrete parameter, can increase the redundancy, depressing the information.
This issue is biologically relevant 
since in several cortical areas neurons show a tuning for the direction, and in particular in the data set analyzed in 
\cite{vale} such correlations were indeed observed. 

Modifying accordingly the theoretical model did improve the fit of real information curves, suggesting that 
the correlations detected in the data are information bearing- here, in a negative sense: they depress the information 
content.

We have been able to study analytically the asymptotic approach to the upper information bound with purely discrete stimuli, always 
without replicas, calculating both the leading term order $\sigma^N$ and the correction order $\sigma^{2N}$.
We have shown how to retrieve our results using a different approach \cite{sompo+00}, always replica free. This approach does allows to 
go beyond our original approximation. Yet, in presence of additional quenched disorder as in the case of our specific model, 
further assumptions 
are necessary to proceed with the analytical evaluation of the correction. 
A careful analysis is still in progress and will be presented elsewhere \cite{vale+03a}.

Finally we have evaluated the asymptotic information value in presence of mixed continuous and discrete stimuli, which grows 
to infinity when the noise goes to zero and/or the number of neurons becomes large. 
We have found that the information grows to infinity logarithmically with $N$ and with the inverse of the noise 
for the gaussian model, while the exact dependence for the thresholded gaussian model is more difficult to 
detect. 
Under certain conditions and for very low values of the noise the asymptotic information is higher for the thresholded gaussian model 
than for the pure gaussian. 

This result is quite interesting per se, but we refrain from any speculation leaving 
its interpretation to a more careful evaluation of the information. 
In particular it will be interesting to evaluate the impact of the corrections neglected in eq.(\ref{info_second_appr}) \cite{vale+03a}.

It must be said that the validity of our results is not checked here by means of numerical simulations.
In a previous work \cite{vale+01c} we have shown and discussed in detail that, since information is a very sensitive measure 
to limited sampling, in our specific case of a continuous rate model with additional quenched parameters,
the numerical evaluation results extremely hard, especially in the most interesting limit of large population sizes.
The simulations presented in \cite{vale+01c} could be carried out using a decoding procedure which is meant to reduce the bias. 
Even so, the agreement with the analytical results was found only for a population size of maximum $2$ cells, the curve deviating 
due to the distortion caused by decoding for larger population sizes.
We are currently working in order to improve the numerical techniques and obtain a better check of our analytical results 
for most of the parameter space.
Nonetheless, we think that the widely established difficulty in getting accurate numerical information estimates with 
models of this type makes our analytical efforts and the results presented here even more remarkable.

\section*{Acknowledgments}
We thank Anthony Coolen for very useful discussions and suggestions. 
Comments from Alessandro Treves, Desire' Bolle' 
and Julie Goulet are acknowledged.
Financial support from EC contract QLGA-CT-2001-51056.

\appendix
\section{The small $\sigma$ limit in presence of purely discrete stimuli}
\label{appA}

Let us reconsider eq.(\ref{small_sigma}). Inserting the expansion (\ref{small_sigma_exp}) one obtains:

\ba
&&I(\{\eta_i\},s)\simeq \log_2 p-\frac{1}{\ln 2}\left\langle \sum_{s}p(s)\left[
\sum_{s^\prime\neq s}\prod_i  e^{-(\eta_i^s-\eta_i^{s^\prime})^2/2\sigma^2}\right]\right\rangle_\eta\nonumber\\
&&\simeq\log_2p-\frac{1}{\ln 2}\left(\sum_{s}p(s)\sum_{k=1}^{\infty} 
\sum_{s_1\neq s}..\sum_{s_k\neq s}
\left\langle\exp\left[-\sum_{l=1}^k\left(\eta^s-\eta^{s_l}\right)^2/2\sigma^2\right]
\right\rangle^N_\eta\right);
\label{full_exp}
\ea
where I have used the fact that quenched disorder is uncorrelated and identically distributed 
across neurons and stimuli.

When $\sigma$ becomes very small one can use the following approximation:
\be
e^{-\left(\eta^s-\eta^{s_1}\right)^2/2\sigma^2}\longrightarrow\sqrt{2\pi\sigma^2} 
\delta\left(\eta^s-\eta^{s_1}\right);
\label{delta_s1}
\ee

Let us reconsider the term with $k=1$ in eq.(\ref{full_exp}):

\ba
&&\sum_{s_1\neq s}\left\langle e^{-\left(\eta^s-\eta^{s_1}\right)^2/2\sigma^2}
\right\rangle_\eta^N\longrightarrow \sum_{s_1\neq s} \left(\sqrt{2\pi}\sigma\right)^N
\left[\int d\eta^s d\eta^{s_1} \varrho(s)\varrho(s_1)\delta\left(\eta^s-\eta^{s_1}\right)\right]^N\nonumber\\
&&=\left(p-1\right)\left(\sqrt{2\pi}\sigma\right)^N
\left[\int d\eta \varrho^2(\eta)\right]^N;
\ea
Therefore this term gives a contribution of order $\sigma^N$. 

It is easy to check that at each order 
$k$ the term with $s_1=s_2..=s_k\neq s$ gives a contribution of order $\sigma^N$: in fact 
while for a generic choice of $s_1$,$s_2$..$s_k$ out of the $p$ correlates one has finally
{\bf several} $\delta$-functions each of which, according to eq.(\ref{delta_s1}), 
carries a factor $\sigma$, when all the stimuli are equal only one 
$\delta$-function remains and the result is of order $\sigma^N$.
Therefore one has to sum all the contributions to calculate the exact coefficient 
determining the asymptotic approach of the information to the upper bound.
For a generic order $k$ in the expansion of the logarithm one has:

\ba
&&\frac{(-1)^{k+1}}{k}\sum_{s_1=s_2..=s_k\neq s}
\left\langle e^{-\sum_{l=1}^k\left(\eta^s-\eta^{s_l}\right)^2/2\sigma^2}
\right\rangle^N_\eta\rightarrow\frac{(-1)^{k+1}}{k}\sum_{s_1\neq s}
\left\langle e^{-k\left(\eta^s-\eta^{s_1}\right)^2/2\sigma^2}\right\rangle_\eta^N\nonumber\\
&&\simeq \frac{(-1)^{k+1}}{k^{N/2+1}}\left(p-1\right)
\left(\sqrt{2\pi}\sigma\right)^N\left[\int d\eta \varrho^2(\eta)\right]^N;
\ea
It is now clear that to obtain the contribution at all orders $k$ one must sum the series:
\be 
S_1=\sum_{k=1}^\infty \frac{(-1)^{k+1}}{k^{N/2+1}};
\ee
and the final contribution to the mutual information up to order $\sigma^N$ can be expressed 
as follows:
\be 
\left(p-1\right)S_1 \left(\sqrt{2\pi}\sigma I_1\right)^N;\,\,\,\,I_1=\int d\eta \varrho^2(\eta);
\label{final_leading}
\ee
Inserting eq.(\ref{final_leading}) in eq.(\ref{full_exp}) one obtains the 
expression of the asymptotic approach to the ceiling up to order $\sigma^N$:
\be
I(\{\eta_i\},s)\simeq \log_2 p\left[1-\frac{p-1}{\log_2p\ln 2}
\left(S_1\left(\sqrt{2\pi}\sigma I_1\right)^N\right)\right];
\label{info_1}
\ee

This result can be easily extended to higher powers of $\sigma$. The first 
correction of order $\sigma^{2N}$ to the leading term can be calculated through the same technique.
Since at the $k^{th}$ order in the expansion of the logarithm the factor $\sigma^N$ 
was obtained considering only the configuration where all the stimuli 
$s_1$..$s_k$ are equal, it is clear that each configuration where all the stimuli except 
one are equal will generate a factor $\sigma^{2N}$; in fact if, say, $l$ stimuli 
are different from one another among the $k$, one will have to introduce a $\delta$-function 
for each of the $l$ exponentials, 
according to eq.(\ref{delta_s1}).
Let us see in detail the $k^{th}$ order contribution, assuming 
for example that all stimuli are equal except $s_1$:

\ba
&&\frac{(-1)^{k+1}}{k}\sum_{s_1\neq s}\sum_{s_2=s_3..=s_k\neq s,s_1}
\left\langle e^{-\sum_{l=1}^k\left(\eta^s-\eta^{s_l}\right)^2/2\sigma^2}
\right\rangle^N_\eta\nonumber\\
&&\rightarrow\frac{(-1)^{k+1}}{k}\sum_{s_1\neq s}\sum_{s_2\neq s_1,s}
\left\langle e^{-\left(\eta^s-\eta^{s_1}\right)^2/2\sigma^2}
e^{-(k-1)\left(\eta^s-\eta^{s_2}\right)^2/2\sigma^2}\right\rangle_\eta^N;
\label{exp_2}
\ea
A $\delta$-function is introduced according to eq.(\ref{delta_s1}) for each
of the two exponentials in eq.(\ref{exp_2}); since one has $k$ possible choices 
for the stimulus which is different from the other $k-1$, the final result must 
be multiplied times a factor $k$ more; finally ne has:
\be
\frac{(-1)^{k+1}}{(k-1)^{N/2}}\left(p-1\right)\left(p-2\right)
\left(2\pi\sigma^2\right)^N\left[\int d\eta \varrho^3(\eta)\right]^N;
\ee
Summing all terms at any order $k$ the total contribution order $\sigma^{2N}$ 
to the mutual information can be written as follows:
\be
\left(p-1\right)\left(p-2\right)S_2
\left(2\pi\sigma^2I_2\right)^N;\,\,\,\,
S_2=\sum_{k=1}^\infty \frac{(-1)^{k}}{k^{N/2}};
\,\,\,\,I_2=\int d\eta \varrho^3(\eta);
\label{correct}
\ee
Inserting this result in eq.(\ref{info_1}) one obtains the final expression for the 
mutual information up to order $\sigma^{2N}$, eq.(\ref{final_info}).  

\section{Large $\sigma$ limit for the thresholded-gaussian model}
\label{appB}

Let us reconsider eqs.(\ref{log_exp_2}). Developing all the products one can 
expand the logarithm in powers of $1/\sigma$:

\ba
&&\log_2 \left[\sum_{s^\prime}p(s^\prime)\int d\vartheta^\prime p(\vartheta^\prime)
\left(1+\sum_{i=1}^k \frac{\eta_i\tilde{\eta}_i(\vartheta^\prime,s^\prime)
}{\sigma^2}-\sum_{i=1}^k\frac{\tilde{\eta}^2_i(\vartheta^\prime,s^\prime)
}{2\sigma^2}+\frac{1}{2}\sum_{i,j=1}^k\frac{
\eta_i\eta_j\tilde{\eta}_i(\vartheta^\prime,s^\prime)
\tilde{\eta}_j(\vartheta^\prime,s^\prime)
}{\sigma^4}\right)\right.\nonumber\\
&&\left. \frac{1}{2^{N-k}}\left(1-\sum_{i=k+1}^{N}\sqrt{\frac{2}{\pi}}
\frac{\tilde{\eta}_i(\vartheta^\prime,s^\prime)}{\sigma}+\frac{1}{\pi}\sum_{i\neq j}
\frac{\tilde{\eta}_i(\vartheta^\prime,s^\prime)
\tilde{\eta}_j(\vartheta^\prime,s^\prime)}{\sigma^2}\right)\right]\nonumber\\
&&\simeq -(N-k)+\frac{1}{\ln 2}\ln\left[1+\sum_{s^\prime}
\int d\vartheta^\prime p(\vartheta^\prime,s^\prime)
\left(\sum_{i=1}^k \frac{\eta_i\tilde{\eta}_i(\vartheta^\prime,s^\prime)
}{\sigma^2}-\sum_{i=1}^k\frac{\tilde{\eta}^2_i(\vartheta^\prime,s^\prime)
}{2\sigma^2}\right.\right.\nonumber\\
&&\left.\left.-\sum_{i=k+1}^{N}\sqrt{\frac{2}{\pi}}
\frac{\tilde{\eta}_i(\vartheta^\prime,s^\prime)}{\sigma}
+\frac{1}{2}\sum_{i,j=1}^k\frac{
\eta_i\eta_j\tilde{\eta}_i(\vartheta^\prime,s^\prime)
\tilde{\eta}_j(\vartheta^\prime,s^\prime)
}{\sigma^4}+\frac{1}{\pi}\sum_{i\neq j}
\frac{\tilde{\eta}_i(\vartheta^\prime,s^\prime)
\tilde{\eta}_j(\vartheta^\prime,s^\prime)}{\sigma^2}\right.\right.\nonumber\\
&&\left.\left.-\sum_{i=1}^{k}\sum_{j=k+1}^{N}
\sqrt{\frac{2}{\pi}}
\frac{\eta_i\tilde{\eta}_i(\vartheta^\prime,s^\prime)
\tilde{\eta}_j(\vartheta^\prime,s^\prime)}{\sigma^3}
\right)\right]\nonumber\\
&&\simeq -(N-k)+\frac{1}{\ln 2}\left[\sum_{s^\prime}
\int d\vartheta^\prime p(\vartheta^\prime,s^\prime)
\left(\sum_{i=1}^k \frac{\eta_i\tilde{\eta}_i(\vartheta^\prime,s^\prime)
}{\sigma^2}-\sum_{i=1}^k\frac{\tilde{\eta}^2_i(\vartheta^\prime,s^\prime)
}{2\sigma^2}\right.\right.\nonumber\\
&&\left.\left.-\sum_{i=k+1}^{N}\sqrt{\frac{2}{\pi}}
\frac{\tilde{\eta}_i(\vartheta^\prime,s^\prime)}{\sigma}
+\frac{1}{2}\sum_{i,j=1}^k\frac{
\eta_i\eta_j\tilde{\eta}_i(\vartheta^\prime,s^\prime)
\tilde{\eta}_j(\vartheta^\prime,s^\prime)
}{\sigma^4}+\frac{1}{\pi}\sum_{i\neq j}
\frac{\tilde{\eta}_i(\vartheta^\prime,s^\prime)
\tilde{\eta}_j(\vartheta^\prime,s^\prime)}{\sigma^2}\right.\right.\nonumber\\
&&\left.\left.-\sum_{i=1}^{k}\sum_{j=k+1}^{N}
\sqrt{\frac{2}{\pi}}
\frac{\eta_i\tilde{\eta}_i(\vartheta^\prime,s^\prime)
\tilde{\eta}_j(\vartheta^\prime,s^\prime)}{\sigma^3}
\right)-\frac{1}{2}\sum_{s^\prime,s^{\prime\prime}}
\int d\vartheta^\prime d\vartheta^{\prime\prime}p(\vartheta^\prime,s^\prime)
p(\vartheta^{\prime\prime},s^{\prime\prime})
\left(\sum_{i,j=1}^k\frac{
\eta_i\eta_j\tilde{\eta}_i(\vartheta^\prime,s^\prime)
\tilde{\eta}_j(\vartheta^{\prime\prime},s^{\prime\prime})
}{\sigma^4}\right.\right.\nonumber\\
&&\left.\left.+\frac{2}{\pi}\sum_{i\neq j}
\frac{\tilde{\eta}_i(\vartheta^\prime,s^\prime)
\tilde{\eta}_j(\vartheta^{\prime\prime},s^{\prime\prime})}{\sigma^2}
-2\sum_{i=1}^{k}\sum_{j=k+1}^{N}
\sqrt{\frac{2}{\pi}}
\frac{\eta_i\tilde{\eta}_i(\vartheta^\prime,s^\prime)
\tilde{\eta}_j(\vartheta^{\prime\prime},s^{\prime\prime})}{\sigma^3}
\right)
\right]
\label{log_exp_3}
\ea
where we have kept only the terms which will give a contribution 
order $1/\sigma^2$ to the information.

This expression has to be inserted in eq.(\ref{info_no_appr}) 
and integrated on $\{\eta_i\}$. One obtains:

\ba
&&-\left\langle\sum_{s}\int d\vartheta p(\vartheta,s)\right.\nonumber\\
&&\left.\sum_{k=0}
^N 
\left(\begin{array}{c}
 N\\k\end{array}
\right)
\int_0^\infty \prod_{i=1}
^k d\eta_i \prod_{i=1}
^k \frac{1}{\sqrt{2\pi\sigma^2}}
exp-\left[\left(\eta_i-\tilde{\eta}_i(\vartheta,s)
\right)^2/2\sigma^2\right]
\prod_{i=k+1}
^N 
(1-\erf(\tilde{\eta}_i(\vartheta,s)/\sigma)) 
\right.\nonumber\\
&&\left.
\log_2\left[\sum_{s^\prime}p(s^\prime)\int d\vartheta^\prime p(\vartheta^\prime)
\left(\prod_{i=1}
^k 
exp\left[2\eta_i\tilde{\eta}_i(\vartheta^\prime,s^\prime)
-\tilde{\eta}_i^2(\vartheta^\prime,s^\prime)/2\sigma^2\right]
\prod_{i=k+1}
^N 
(1-\erf(\tilde{\eta}_i(\vartheta^\prime,s^\prime)/\sigma)) 
\right)\right]\right\rangle_{\varepsilon,\vartheta^0}\nonumber\\
&&\simeq
\sum_{s,s^\prime,s^{\prime\prime}}\int d\vartheta d\vartheta^\prime d\vartheta^{\prime\prime}
p(s,\vartheta)p(s^\prime,\vartheta^\prime)p(s^{\prime\prime},\vartheta^{\prime\prime})\nonumber\\
&&\left[\sum_{k=0}
^{N-1} 
\left(\begin{array}{c}N\\k\end{array}
\right)
\left(N-k\right)\left\langle 
\erf(\tilde{\eta}(\vartheta,s)/\sigma)\right\rangle^k_{\varepsilon,\vartheta^0}
\left\langle 
1-\erf(\tilde{\eta}(\vartheta,s)/\sigma)\right\rangle^{N-k}_{\varepsilon,\vartheta^0}
\right.\nonumber\\
&&\left.-\sum_{k=1}
^{N} 
\left(\begin{array}{c}
N\\k\end{array}
\right)
\frac{k}{\ln 2}\left\langle \tilde{\eta}(\vartheta,s)
\tilde{\eta}(\vartheta^\prime,s^\prime)/\sigma^2
\erf(\tilde{\eta}(\vartheta,s)/\sigma)\right\rangle_{\varepsilon,\vartheta^0}
\left\langle 
\erf(\tilde{\eta}(\vartheta,s)/\sigma)\right\rangle^{k-1}_{\varepsilon,\vartheta^0}
\left\langle 
1-\erf(\tilde{\eta}(\vartheta,s)/\sigma)\right\rangle^{N-k}_{\varepsilon,\vartheta^0}\right.\nonumber\\
&&\left.-\sum_{k=1}
^N 
\left(\begin{array}{c}
 N\\k\end{array}
\right)
\frac{k}{\ln 2} \left\langle \tilde{\eta}(\vartheta^\prime,s^\prime)/\sqrt{2\pi}\sigma
e^{-\tilde{\eta}^2(\vartheta,s)/2\sigma^2}\right\rangle_{\varepsilon,\vartheta^0}
\left\langle 
\erf(\tilde{\eta}(\vartheta,s)/\sigma)\right\rangle^{k-1}_{\varepsilon,\vartheta^0}
\left\langle 
1-\erf(\tilde{\eta}(\vartheta,s)/\sigma)\right\rangle^{N-k}_{\varepsilon,\vartheta^0}\right.\nonumber\\
&&\left.+\sum_{k=0}
^{N-1} 
\left(\begin{array}{c}
N\\k\end{array}
\right)
\frac{N-k}{\ln 2}\sqrt{\frac{2}{\pi}}
\left\langle \frac{\tilde{\eta}(\vartheta^\prime,s^\prime)}{\sigma}
[1-\erf(\tilde{\eta}(\vartheta,s)/\sigma)]\right\rangle_{\varepsilon,\vartheta^0}
\left\langle 
\erf(\tilde{\eta}(\vartheta,s)/\sigma)\right\rangle^{k}_{\varepsilon,\vartheta^0}
\left\langle 
1-\erf(\tilde{\eta}(\vartheta,s)/\sigma)\right\rangle^{N-k-1}_{\varepsilon,\vartheta^0}\right.\nonumber\\
&&\left.+\sum_{k=1}
^{N} 
\left(\begin{array}{c}
N\\k\end{array}
\right)
\frac{k}{2\ln 2}\left\langle \frac{\tilde{\eta}(\vartheta^\prime,s^\prime)
\tilde{\eta}(\vartheta^{\prime\prime},s^{\prime\prime})}{\sigma^2}
\erf(\tilde{\eta}(\vartheta,s)/\sigma)\right\rangle_{\varepsilon,\vartheta^0}
\left\langle 
\erf(\tilde{\eta}(\vartheta,s)/\sigma)\right\rangle^{k-1}_{\varepsilon,\vartheta^0}
\left\langle 
1-\erf(\tilde{\eta}(\vartheta,s)/\sigma)\right\rangle^{N-k}_{\varepsilon,\vartheta^0}\right.\nonumber\\
&&\left.+\sum_{k=0}
^{N-1} 
\left(\begin{array}{c}
N\\k\end{array}
\right)
\frac{N-k}{\pi\ln 2}
\left\langle \tilde{\eta}(\vartheta^\prime,s^\prime)
\tilde{\eta}(\vartheta^{\prime\prime},s^{\prime\prime})/\sigma^2
[1-\erf(\tilde{\eta}(\vartheta,s)/\sigma)]\right\rangle_{\varepsilon,\vartheta^0}\right.\nonumber\\
&&\left.
\left\langle 
\erf(\tilde{\eta}(\vartheta,s)/\sigma)\right\rangle^{k}_{\varepsilon,\vartheta^0}
\left\langle 
1-\erf(\tilde{\eta}(\vartheta,s)/\sigma)\right\rangle^{N-k-1}_{\varepsilon,\vartheta^0}
\right]
\label{log_k}
\ea
where we have assumed that quenched disorder is uncorrelated across neurons.

It must be noticed that the terms of order $N^2$ appearing in the expansion of the logarithm, 
eq.(\ref{log_exp_3}) do not appear in eq.(\ref{log_k}) any more: it can be easily shown that 
after averaging across the stimuli they cancel out.

The sums on $k$ in eq.(\ref{log_k}) can be performed and the result is then inserted in 
the expression for the mutual information, eq.(\ref{info_no_appr}), which finally can 
be rewritten as follows:

\ba
&&I(\{\eta_i\},\vartheta\otimes s)=\sum_s\int d\vartheta p(\vartheta,s)
\left[\frac{N}{\ln 2}\left\langle\left[1-\erf(\tilde{\eta}(\vartheta,s)/\sigma)\right]
\ln\left[1-\erf(\tilde{\eta}(\vartheta,s)/\sigma)\right]
\right\rangle_{\varepsilon,\vartheta^0}\right.\nonumber\\
&&\left.+\frac{N}{2\ln 2}
\left\langle\frac{\tilde{\eta}^2(\vartheta,s)}{\sigma^2}
\erf(\tilde{\eta}(\vartheta,s)/\sigma)
\right\rangle_{\varepsilon,\vartheta^0}
+N\left\langle\left[1-\erf(\tilde{\eta}(\vartheta,s)/\sigma)\right]
\right\rangle_{\varepsilon,\vartheta^0}
\right]\nonumber\\
&&-\sum_{s,s^\prime,s^{\prime\prime}}
\int d\vartheta d\vartheta^\prime d\vartheta^{\prime\prime}
p(s,\vartheta)p(s^\prime,\vartheta^\prime)p(s^{\prime\prime},\vartheta^{\prime\prime})
\nonumber\\
&&\left[\frac{N}{\ln 2}
\left\langle\frac{\tilde{\eta}(\vartheta,s)\tilde{\eta}(\vartheta^\prime,s^\prime)}{\sigma^2}
\erf(\tilde{\eta}(\vartheta,s)/\sigma)
\right\rangle_{\varepsilon,\vartheta^0}-\frac{N}{2\ln 2}
\left\langle\frac{\tilde{\eta}(\vartheta^\prime,s^\prime)
\tilde{\eta}(\vartheta^{\prime\prime},s^{\prime\prime})}{\sigma^2}
\erf(\tilde{\eta}(\vartheta,s)/\sigma)
\right\rangle_{\varepsilon,\vartheta^0}\right.\nonumber\\
&&\left.-\frac{N}{\ln 2}\sqrt{\frac{2}{\pi}}
\left\langle\frac{\tilde{\eta}(\vartheta^\prime,s^\prime)}{\sigma}
[1-\erf(\tilde{\eta}(\vartheta,s)/\sigma)]
\right\rangle_{\varepsilon,\vartheta^0}
-\frac{N}{\pi\ln 2}
\left\langle\frac{\tilde{\eta}(\vartheta^\prime,s^\prime)
\tilde{\eta}(\vartheta^{\prime\prime},s^{\prime\prime})}{\sigma^2}
[1-\erf(\tilde{\eta}(\vartheta,s)/\sigma)]
\right\rangle_{\varepsilon,\vartheta^0}\right]
\label{pre_info}
\ea
Using the expansion (\ref{erf_exp}) for the error function and keeping only the terms 
up to order $N/\sigma^2$ one arrives at eq.(\ref{info_large_corr}).


\end{document}